\documentclass{ejm}
\usepackage{latexsym,epsfig,amssymb,amsmath,subfigure,color,pictex}
\usepackage{hyperref}
\usepackage{multirow}
\usepackage{psfrag}
\usepackage{epstopdf}
\usepackage{amsmath}
\usepackage{afterpage} 
\usepackage{makecell}
\usepackage{cleveref}
\usepackage{natbib}

\ifprodtf \else
  \checkfont{eurm10}
  \iffontfound
    \IfFileExists{upmath.sty}
      {\typeout{^^JFound AMS Euler Roman fonts on the system,
                   using the 'upmath' package.^^J}%
       \usepackage{upmath}}
      {\typeout{^^JFound AMS Euler Roman fonts on the system, but you
                   don't seem to have the}%
       \typeout{'upmath' package installed. EJM.cls can take advantage
                 of these fonts,^^Jif you use 'upmath' package.^^J}%
      }
  \else
  \fi
\fi

\ifprodtf \else
  \checkfont{msam10}
  \iffontfound
    \IfFileExists{amssymb.sty}
      {\typeout{^^JFound AMS Symbol fonts on the system, using the
                'amssymb' package.^^J}%
       \usepackage{amssymb}%
         \let\leq=\leqslant
         
      }{}
  \fi
\fi

\ifprodtf \else
  \IfFileExists{amsbsy.sty}
 
    {\typeout{^^JFound the 'amsbsy' package on the system, using it.^^J}%
     \usepackage{amsbsy}}
    {}
\fi

\graphicspath{{./}{./figures/}}

\newcommand{\bes} {\begin{eqnarray*}}
\newcommand{\ees} {\end{eqnarray*}}
\newcommand{\ds}{\displaystyle}
\newcommand{\be}{\begin{eqnarray}}
\newcommand{\ee}{\end{eqnarray}}
\newcommand{\vect}[1]{\mbox{\boldmath $#1$}}

\newcommand{\dd} \partial
\newcommand{\vep}{\varepsilon}

\newcommand{\non} \nonumber

\newcommand{\mub}{\bar{\mu}}

\newcommand{\tn}{t_{+}^{0}}
\newcommand{\de}{{D_{\rm eff}}}
\newcommand{\dc}{{\cal D}}
\newcommand{\src}{{\cal S}}

\newcommand{\jj}{\vect{j}}
\newcommand{\dha}{\vect{\hat{d}}}

\newcommand{\xx}{\vect{x}}
\newcommand{\vp}{\vect{v}_p}
\newcommand{\vn}{\vect{v}_n}

\newcommand{\vph}{\varphi}
\newcommand{\x}{\vect{x}}
\newcommand{\q}{\vect{q}}
\newcommand{\qn}{\vect{q}_n}
\newcommand{\qp}{\vect{q}_p}

\newcommand{\ev}{\epsilon_v}
\newcommand{\bet}{b_{et}}
\newcommand{\B}{\underline{\underline{B}}}

\newcommand{\VP}{\hat{V}_{\mathrm{per}}}
\newcommand{\OP}{\hat{\Omega}_{\mathrm{per}}}
\newcommand{\dV}{{\rm d} \hat{V}}
\newcommand{\dS}{{\rm d} \hat{S}}

\newcommand{\mv}{|\hat{V}_{\mathrm{per}}|}
\newcommand{\mo}{|\hat{\Omega}_{\mathrm{per}}|}
\newcommand{\so}{|S_{\dd \OP}|}

\newcommand{\Vnh}{{V}_{\mathrm{per}}}
\newcommand{\OPnh}{{\Omega}_{\mathrm{per}}}

\newcommand{\avj}{\langle \jj \rangle}
\newcommand{\avsj}{\langle j \rangle}

\newcommand{\avqp}{\langle \vect{q}_p \rangle}
\newcommand{\avjn}{\bar{j}_{\rm tr}}

\newcommand{\jsa}{j_{\rm s}^{(\rm a)}}
\newcommand{\jsc}{j_{\rm s}^{(\rm c)}}
\newcommand{\dsa}{D_{\rm s}^{(\rm a)}}
\newcommand{\dsc}{D_{\rm s}^{(\rm c)}}
\newcommand{\jtra}{j_{\rm tr}^{(\rm a)}}
\newcommand{\jtrc}{j_{\rm tr}^{(\rm c)}}
\newcommand{\csa}{c_{\rm s}^{(\rm a)}}
\newcommand{\cscc}{c_{\rm s}^{(\rm c)}}
\newcommand{\phisa}{\phi_{\rm s}^{(\rm a)}}
\newcommand{\phisc}{\phi_{\rm s}^{(\rm c)}}
\newcommand{\Ra}{R^{(\rm a)}}
\newcommand{\Rc}{R^{(\rm c)}}
\newcommand{\bta}{b_{\rm et}^{(\rm a)}}
\newcommand{\btc}{b_{\rm et}^{(\rm c)}}
\newcommand{\kapsa}{\kappa_{\rm s}^{(\rm a)}}
\newcommand{\kapsc}{\kappa_{\rm s}^{(\rm c)}}

\newcommand{\nn}{\vect{n}}

\newcommand{\ie}{{\it i.e.}\ }
\newcommand{\eg}{{\it e.g.}\ }

\newcommand{\s}{c_0}
\newcommand{\p}{c_p}
\newcommand{\n}{c_n}

\title{Charge transport modelling of lithium ion batteries}
\author[G. W. Richardson et al.]{%
G.\ns W.\ns R\ls I\ls C\ls H\ls A\ls R\ls D\ls S\ls O\ls N$\,^{1,2}$\ns,
J.\ns M. \ns F\ls O\ls S\ls T\ls E\ls R$\,^{2,4}$\ns,
R.\ns  R\ls A\ls N\ls O\ls M$\,^5$\ns,
C.\ns P.\ns P\ls L\ls E\ls A\ls S\ls E$\,^{2,6}$\ns
\and \ns
A.\ns M.\ns R\ls A\ls M\ls O\ls S$\,^3$
}
\affiliation{%
$^1\,$
School of Mathematics, University of Southampton, Southampton SO17 1BJ, UK\\
email\textup{\nocorr: \texttt{g.richardson@soton.ac.uk}}\\
$^2\,$
The Faraday Institution, Quad One, Becquerel Avenue, Harwell Campus, Didcot, OX11 0RA, UK\\
$^3\,$
{Instituto de Matem\'atica Interdisciplinar \& Departamento de An\'alisis Matem\'atico y} Matem\'atica Aplicada, Universidad Complutense de Madrid, Plaxa de Ciencias, 3, 28040 Madrid, Spain\\
$^4\,$
School of Mathematics and Physics, University of Portsmouth, Portsmouth, PO1 2UP.\\
$^5\,$
Faculty of Electrical Engineering, Universiti Teknikal Malaysia Melaka,
Hang Tuah Jaya, 76100 Durian Tunggal, Melaka, Malaysia.\\
$^6\,$
Mathematical Institute, University of Oxford, Woodstock Rd, Oxford OX2 6GG}
\date{10 December 2019}
\pubyear{2017}
\volume{000}
\pagerange{\pageref{firstpage}--\pageref{lastpage}}

\begin{document}

\maketitle

\begin{abstract}
  This paper presents the current state of mathematical modelling of
  the electrochemical behaviour of lithium-ion batteries as they are
  charged and discharged. It reviews the models developed by Newman
  and co-workers, both in the cases of dilute and
  moderately-concentrated electrolytes and indicates the modelling
  assumptions required for their development. Particular attention is
  paid to the interface conditions imposed between the electrolyte and
  the active electrode material; necessary conditions are derived for
  one of these, the Butler-Volmer relation, in order to ensure
  physically realistic solutions. Insight into the origin of the differences between
  various models found in the literature is revealed by considering
  formulations obtained by using different measures of the electric
  potential. Materials commonly used for electrodes in lithium ion
  batteries are considered and the various mathematical models used to
  describe lithium transport in them discussed. The problem of
  up-scaling from models of behaviour at the single electrode particle
  scale to the cell scale is addressed using homogenisation techniques
  resulting in the pseudo 2D model commonly used to describe
 charge transport and discharge behaviour in lithium-ion cells. Numerical solution to this model is discussed and illustrative results for a common device are computed.
\end{abstract}
\label{firstpage}



\section{Introduction \label{1}}
Lithium-ion batteries are currently one of the most hopeful prospects
for large scale efficient storage of electricity for mobile devices
from phones to cars. Crucial to their continued improved performance
is to understand how novel materials might be effectively exploited in
their design. Excellent reviews of the current status of such
materials are given by \cite{bruce08}, \cite{choi12}, {\cite{blom17}}. Understanding how these materials affect macroscopic
battery behaviour is greatly aided by good mathematical models of the
transport processes within the battery.

The purpose of this paper is to serve as a guide to charge transport
modelling in lithium-ion batteries. Much of the work in this area is
due to John Newman and his co-workers who, in a series of seminal
publications \cite{1973Ne,newman-book,doyle93,doyle96,fuller94,ma95,newman03,srinivasan04},
introduced and applied models for these devices that account both
for charge transport in the electrolyte as well as solid lithium
ion diffusion in the active electrode materials, and use Butler-Volmer
reaction kinetics for lithium intercalation and de-intercalation on
the electrode/electrolyte surface to couple these two processes
together. While these models have been remarkably successful in
describing the behaviour of real batteries they are not easily
extracted from the literature and, in addition, improvements in understanding of electrode materials have led to significant recent advances in the modelling of lithium transport in electrode particles that can incorporated into this framework. This work aims to  provide a relatively concise guide to the subject
while at the same time highlighting some of the common modelling
pitfalls.

\begin{figure}
\begin{center}
\includegraphics[width=\columnwidth]{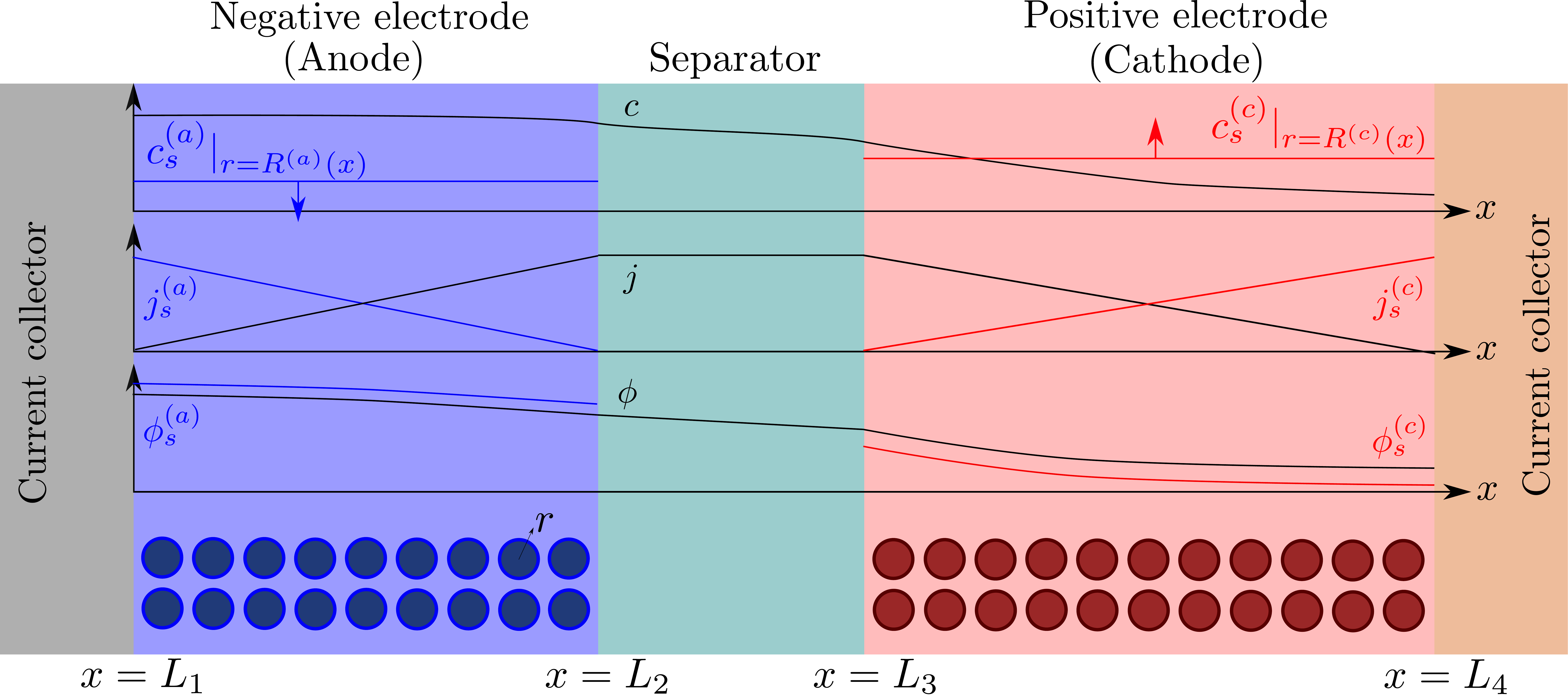}
\caption{A sketch of a cross section of a typical device as well as the macroscopic variables and their domains of definition. }
\label{geom_sketch}
\end{center}
\end{figure}

A typical lithium-ion battery (LIB) cell has three regions: i) a porous negative
electrode, ii) a porous positive electrode and iii) an
electron-blocking separator (see figure \ref{geom_sketch}). Typically the electrodes are comprised of
particles, typically a few microns in size, of {\it different} (solid)
active materials (AMs), that are capable of absorbing lithium into
their structure and therefore act as lithium reservoirs. These particles are interspersed
with an inert porous polymer binder material, combined with highly conducting
carbon black, that acts to hold the electrode particles in place and
form conducting links between electrode particles. The AM particle and polymer binder regions
are permeated by a lithium electrolyte that serves to transport charge
and lithium ions between the AMs of the two electrodes, with direct
electrical contact between the negative and positive electrodes being
prevented by the presence of the porous separator.  On the interfaces
between the electrode particles and the electrolyte (de-)intercalation
reactions, in which lithium ions transfer between the electrolyte and
the AM, take place. The AM of the positive electrode shows a greater
affinity for lithium than that of the negative electrode so that in a
charged cell there is a propensity for a current of positively charged
lithium ions to flow from the negative to the positive electrode and
thereby establishing a useful potential difference between the
electrodes. The rates at which these (de-)intercalation reactions
take place on the electrode/electrolyte interfaces are key to the
electrical behaviour of the battery, and are typically described by a
Butler-Volmer relation which gives the interfacial current density,
from AM to electrolyte, in terms of the potential jump between the AM
and the electrolyte and the lithium concentrations in the AM and the
electrolyte.

In addition to discussing transport models we also propose a new
formulation for the Butler-Volmer relation, for current flow between the
AM and electrolyte, with the aim of ensuring a model that is able to
simulate extreme cases, where classical formulations lead to
physically unrealistic lithium distributions. More precisely, we
address the issues of the limiting conditions in which the electrodes
are close to being fully intercalated (or fully depleted) or in which
the electrolyte concentration is very low.

The LIB is great examplar of multiscale and multiphysics systems. Accurately predicting battery pack (formed from many individual cells) behaviour depends on having appropriate represenatations of the physics and chemistry at a range of smaller scales, including that of individual cells within a pack, individual electrodes within each cell, individual particles within electrodes and at the level of the atomic structure of the materials making up the particles. The behaviour at many of these lengthscales is dictated by a myriad of interacting phenomena including electrochemical ones, but also thermal and mechanical ones. The seminal models developed by Newman and his co-workers span the scales of individual electrodes particles upto individual cells and are largely focussed on the electrochemical behaviour. Many authors including \cite{ranom14,schmuck17} and \cite{franco13} discuss the challenges associated with formulating mathematical models that couple phenomena pccuring at vastly disparate lengthscales. There is also current impetus to experimentally characerise and develop new modelling approaches to better understand chemical structure within electrode materials \cite{harris17} and \cite{kim04}, mechanical and thermal effects \cite{Oh14}, chemical degradation \cite{Seth19,Wang18,birkl17,mon03} and \cite{Nish11}, and to also develop battery management systems to optimally control batteries for use in electric vehicles \cite{Lu13} and \cite{Kim14}.

The outline of this work is as follows. In \S \ref{electrolyte} we
discuss charge transport models for the electrolyte. We begin, in \S \ref{dil}, with the
simplest description, dilute electrolyte theory,  and show that it
cannot adequately describe electrolyte data at the typical
concentrations encountered in real batteries. This motivates us to
consider Newman's moderately-concentrated electrolyte theory
\cite{1973Ne,newman-book} in \S \ref{conc}, which forms the basis for much of
the battery electrolyte modelling currently being undertaken and to
show how this fits to real data for the common electrolyte LiPF$_6$ \cite{valoen05}. In \S
\ref{particles} lithium transport within AM electrode particles is
briefly reviewed while in \S \ref{sdbve} a new formulation for
the Butler-Volmer relation is proposed. In \S \ref{device} the
various strands of the battery chemistry, described in the previous
sections, are brought together to formulate a macroscopic device scale
model for an entire cell. This is accomplished via homogenisation method set out in \S \ref{homog}. In \S\ref{exres} we present a selection of solutions of the device scale model for a common modern device configuration. Finally in \S
\ref{conclusions} we review the key insights of this work.


\section{Modelling the electrolyte \label{electrolyte}}
Here we begin, in \S \ref{dil}, by considering the theory of
very dilute electrolytes, often termed Poisson-Nernst-Planck (PNP) theory.
Because the theory is commonly encountered in modelling semiconductors, is relatively straightforward and
physically appealing, it is useful to highlight some of the
peculiarities associated with charge transport modelling in
batteries. These include, charge neutrality, the use of
electrochemical potentials, and the measurement of the electric
potential with respect to a lithium reference electrode. However the
price of simplicity is that dilute theory does not describe battery
electrolyte behaviour particularly well. Many of these limitations are overcome
in Newman's theory of moderately-concentrated electrolytes, which we
review in \S \ref{conc}. This theory is considerably more involved
than the dilute theory and in practice requires that various functions
be fitted to data directly measured from the electrolyte under
consideration. However, within these limitations, it does provide a
good description of most electrolytes formed by dissolution of a salt
in a solvent. We also hope that by introducing the peculiarities of
notation associated with battery electrolyte modelling in the context
of the simpler dilute theory it will make it easier for the reader to
follow the more complex moderately-concentrated theory.


\subsection{Dilute electrolytes \label{dil}} 
We consider an electrolyte composed of a solvent, a negative
  ion with molar concentration $c_n$ and charge $z_n e$, and a
positive positive ion with molar concentration $c_p$ and charge
  $z_pe$ (where $e$ is the elementary charge and $z_n$, $z_p$ are
  integers accounting for the charge state). This general binary
  electrolyte can be easily studied but because the purpose of this
  section is to give a simple introduction in the rest of this article
  we focus on a 1:1 electrolyte with a generic negative ion and a
  positive lithium ion Li$^+$, so that $z_n=-1$, $z_p=1$.

  Because of the long timescales over which batteries are typically
  charged and discharged it is entirely reasonable to neglect magnetic
  effects and assume that the electric field $\vect{E}$ is
  irrotational (\ie $\nabla \times \vect{E}=\vect{0}$) so that it can
  be written in terms of an electric potential $\phi$, via the
  relation
\be 
\vect{E}=-\nabla \phi.
\label{dil-potential}
\ee
Considering the charge within the system then gives Poisson's equation
\be 
\nabla \cdot (\vep \nabla \phi ) = F (c_n-c_p)\textcolor{red}{,}
\label{dil-poiss}
\ee 
where $\vep$ is the permittivity of the electrolyte.

Since the ions in battery electrolytes do not react with each other or with the solvent, conservation of the two ion species implies that 
\be 
\frac{\dd c_n}{\dd t} + \nabla \cdot \qn=0, \quad \mbox{and}
\quad \frac{\dd c_p}{\dd t} + \nabla \cdot \qp=0\textcolor{red}{,}
\label{dil-cons} 
\ee 
where $\qp$ and $\qn$ are the fluxes of positive and negative ions,
respectively. We can also write this using $\qp=c_p \vect{v}_p$ and
$q_n = c_n \vect{v}_n$ where $\vect{v}_p$ and $\vect{v}_n$ are the
average velocities of the respective species.  In the dilute limit the
electrolyte solvent is assumed to be stationary and the ions to move
in response to thermal diffusion and electric fields; interactions
between ions are neglected.  The component of the average velocity of
a lithium ion due to the electric field, $\vect{v}_{ep}$, is given by
balancing the force $e \vect{E}$, exerted on it by
the electric field, with the viscous drag force $ \vect{v}_{ep}/M_p$,
exerted on it by the solvent (here $M_p$ is the mobility of the
lithium ion).  Thus the advective lithium ion velocity due to the
electric field is $\vect{v}_{ep}=-e M_p \nabla \phi$ and in a similar
manner the average negative ion velocity can be shown to be
$\vect{v}_{en}=e M_n \nabla \phi$.  In addition to the advective
fluxes ($c_p \vect{v}_{ep}$ and $c_n \vect{v}_{en}$) both ion species
diffuse, in response to random thermal excitations. This gives rise to
Fickian fluxes (for positive and negative ions) of size
$-D_p \nabla c_p$ and $-D_n \nabla c_n$, respectively, where $D_p$
and $D_n$ are the respective diffusion coefficients. To highlight the
difference between these diffusion coefficients and those used in the Stefan-Maxwell theory that we will review in \S\ref{conc}, let us point out that $D_p$ (respectively, $D_n$) is the diffusion coefficient for positive (respectivley, negative) ions in a mixture of solvent and negative (respectively, positive) ions. The total ion fluxes, $\qn$ and $\qp$, are obtained by summing their advective and diffusive components, so that
\be 
\qn = c_n \vect{v}_n = -D_n \left( \nabla c_n
  - \frac{e}{k T} c_n \nabla \phi \right), \\
\qp = c_p \vect{v}_p = -D_p\left( \nabla c_p + \frac{e}{k T} c_p \nabla \phi
\right).  
\label{vec-flux}
\ee
Here we have substituted for ion mobilities in terms
of the diffusion coefficients by using the Einstein relations
$M_p=D_p/kT$ and $M_n=D_n/kT$ (where $k$ is Boltzmann's constant). Since
this theory is applied in a chemical setting it is more usual
to write these equations in terms of Faraday's constant $F$ and the
universal gas constant $R$ which, on noting that $e/k=F/R$, leads to the alternative expressions
\be 
\qn= -D_n \left( \nabla c_n - \frac{F}{R T} c_n \nabla \phi
\right), \quad \mbox{and} \quad \qp=-D_p\left( \nabla c_p + \frac{F}{R
    T}c_p \nabla \phi \right).
\label{dil-fluxes} 
\ee 
The equations governing the three variables,
 $c_n$, $c_p$ and $\phi$, are thus (\ref{dil-poiss}),
(\ref{dil-cons}) and (\ref{dil-fluxes}).

\subsubsection{Double layers and charge neutrality.} 
It has long been recognised that (see \eg \cite{newman-book}), at the
concentrations typically encountered in practical electrolytes, there
is almost exact charge neutrality. This implies that there is a balance
between the concentrations of positive and negative charges,
throughout the vast majority of the electrolyte. The exception to this
rule is in the so-called {\it double layers} which lie along the
boundaries of the electrolyte region and are typically extremely
thin, with widths less than a few nanometres. This observation can be
justified mathematically by non-dimensionalising equations (\ref{dil-poiss}),  
(\ref{dil-cons}) and (\ref{dil-fluxes})
and conducting a boundary layer analysis in terms of the small
dimensionless parameter which measures the ratio of the Debye length
(\ie the typical width of a double layer) to the typical dimension of
the electrolyte.\footnote{In fact these equations are unlikely to hold
  inside the double layers but the same procedure can be conducted on
  a generalised version that includes the necessary physics in
  these regions.} The result of such an analysis (see, for example,
\cite{richardson09}) is that, with the exception of the double layers,
$c_n$ must be almost exactly equal to $c_p$. The
  physical meaning of this fact is that the attraction between charges
  is very strong compared to any space charge that the electric field may create. Therefore, a very good approximation to (\ref{dil-poiss}) is 
\be
0= c_n - c_p,
\label{dil-poiss_1} 
\ee 
which is usually called the charge neutrality condition.
As we shall discuss later, because (\ref{dil-poiss_1}) has neglected the derivatives 
that were in (\ref{dil-poiss}), the model needs fewer
boundary conditions at the edges of the
electrolyte (\ie only two boundary conditions are required on the
electrolyte `surface' rather than the three
needed for the full system). 

\subsubsection{The approximate equations.} 
In line with the discussion above we introduce a single concentration $c$ by taking
\be
c_n=c_p=c, \label{electroneut-dil}
\ee
and substitute this into (\ref{dil-cons}) and (\ref{dil-fluxes}) to obtain the approximate charge-neutral equations
\be
\frac{\dd c}{\dd t} + \nabla \cdot \qn=0, \quad \mbox{and} \quad  \qn= -D_n \left( \nabla c - \frac{F}{R T} c \nabla \phi \right), \label{cn1}\\
\frac{\dd c}{\dd t} + \nabla \cdot \qp=0, \quad \mbox{and} \quad \qp=-D_p\left( \nabla c + \frac{F}{R T}c \nabla \phi \right). \label{cn2}
\ee
A common approach to studying this problem is to 
assume that the ionic diffusivities $D_n$ and $D_p$ are constant and rewrite the system by adding (\ref{cn1}a) multiplied by $D_p$ to (\ref{cn2}a) multiplied by $D_n$. On substituting for $\qn$ and $\qp$ this yields a diffusion equation for $c$, of the form
\be
\frac{\dd c}{\dd t} - \de\nabla^2 c = 0 
\quad \mbox{with} \quad  \de=2\frac{D_n D_p}{D_n+D_p}\textcolor{red}{,} 
\label{cn3}
\ee
where $\de$ is termed the effective ionic diffusivity.

Useful physical insight can be found using an alternative formulation of
(\ref{cn1}) and (\ref{cn2}) by introducing
the electric current density, $\jj$, defined in terms of the ion fluxes by
\be
\jj=F(\qp-\qn).
\label{curr}
\ee
Using this concept, a version of Ohm's Law may be obtained by subtracting (\ref{cn1}b) from (\ref{cn2}b), while a charge conservation equation may be found by subtracting (\ref{cn1}a) from (\ref{cn2}a). These may be written in the form
\be
{\jj=-\hat{\kappa}(c) \left[\nabla \phi - \frac{RT}{F} (1- 2t_+) \frac{\nabla c}{c} \right],} 
\label{cn4}
\\
{\nabla \cdot \jj=0,}
\label{cn4-div}
\\
 \mbox{where} \quad \hat{\kappa}(c)=\frac{F^2}{R T}(D_n+D_p) c  \quad \mbox{and} \quad t_+ =\frac{D_p}{D_n+D_p}.
\label{cn4b}
\ee
Here $t_+$ is referred to as the transference number and $\hat{\kappa}(c)$  is referred to as the electrical conductivity of the electrolyte ({\it c.f.} the standard form of Ohm's law is $\jj=-\kappa \nabla\phi$). We use the notation $\hat{\kappa}$ to distinguish the electrical conductivity in the dilute limit from the same quantity in moderately concentrated theory, see (\ref{eq:trans}). The alternative formulation of (\ref{cn1}) and (\ref{cn2}) mentioned above is then (\ref{cn3}), (\ref{cn4}) and (\ref{cn4-div}).

Assuming that the ionic diffusivities $D_n$ and
$D_p$ are constant implies (i) that transference number $t_+$ is
constant, (ii) the effective ionic diffusivity $\de$ is constant and (iii)
electrolyte conductivity $\kappa$ grows linearly with electrolyte
concentration $c$. All three of these quantities are readily measured
experimentally for real electrolytes. For most electrolytes
transference number is usually found to remain close to constant (with
the exception of some polymer electrolytes \eg
\cite{doeff00,fauteux88}), electrolyte diffusivity usually decreases
relatively weakly with concentration, except at very dilute
concentrations, but the growth of electrical conductivity with concentration is far
from linear. Examples of the experimentally measured concentration
dependence of $\de$ and $\kappa$ (from \cite{valoen05}) are
plotted in Figure \ref{fitting} for the battery electrolyte
LiPF$_6$.
Notably
at the typical concentrations used in batteries (roughly 1 molar for
LiPF$_6$) electrical conductivity is nearly constant and 
often lies close to its maximum value, and is
thus not well-approximated by the linear expression in (\ref{cn4b}). 
The explanation given for this poor fit is usually
that even at relatively dilute concentrations there
is a significant drag between ions of opposite charge. The reason for
this is that two ions of opposite charge that lie close to each other
experience a significant electrostatic attraction that negates, to a
large extent, the effects of the global electric field which is trying
to drive the ions in opposite directions.  This observation has
motivated Newman \cite{newman-book} to use Stefan-Maxwell theory for a
multi-component solution to describe the charge transport behaviour of
electrolytes. A summary of the modelling assumptions made in this
theory is given in \S \ref{conc}.

\subsubsection{The dilute theory in terms of electrochemical potentials.}
The dilute model (\ref{cn1})-(\ref{cn2}) can also be written in terms
of the electrochemical potentials, $\mu_n$ and $\mu_p$, of negative
and positive ions respectively. This is the preferred notation for the
ion conservation equations in the electrochemical literature. For an
electrolyte that is formed by ideal salt solution the electrochemical
potentials are given by 
\be 
\mu_n=\mu_n^0+ RT \log \left(
  \frac{c}{c_T} \right) - F \phi, \qquad \mu_p=\mu_p^0+ RT \log \left(
  \frac{c}{c_T} \right) +F \phi, 
\label{ecpots} 
\ee 
where $c_T$ is the
total molar concentration of the electrolyte and, for a dilute
solution, is approximately equal to the solvent concentration. The
first term on the right-hand sides of both expressions in
(\ref{ecpots}) is the standard state potential (per mole of the
species), while the second is the entropy of mixing (per mole of the
species) and the final term is the electrostatic potential (per mole
of the species). Using this notation the conservation equations
(\ref{cn1}) and (\ref{cn2}) can be written in the form 
\be
\frac{\dd c}{\dd t} + \nabla \cdot (c \vn)=0, \quad \mbox{and} \quad  
\vn= -\frac{D_n}{RT} \nabla \mu_n, 
\label{cn5}\\
\frac{\dd c}{\dd t} + \nabla \cdot (c \vp)=0, \quad \mbox{and} \quad
\vp=-\frac{D_p}{RT} \nabla \mu_p. 
\label{cn6} 
\ee 
Here the average
velocities of the two species, $\vn$ and $\vp$, are obtained by
multiplying the gradient of the electrochemical potentials by the
species mobilities, ${D_n}/{RT}$ and ${D_p}/{RT}$. This formalism
extends to nonideal salt solutions and to multicomponent systems. In \S\ref{conc} this approach of using electrochemical potentials is extended to moderately-concentrated electrolytes.

\subsubsection{The potential measured with respect to a lithium electrode.}
The dilute theory as formulated above is at odds with the electrolyte theory used by \cite{newman-book}; here, the factor in front of the concentration gradient in (\ref{cn4}), the constitutive law for the current, is $2({RT}/{F}) (1- t_+)$ rather than $({RT}/{F}) (1- 2 t_+)$ as above. As pointed out in \cite{ramos16}, this has generated some confusion in the literature. The explanation for this discrepancy (as initially demonstrated by \cite{ranom14} and subsequently in \cite{bizeray16}) is that the theory in \cite{newman-book} is formulated in terms of $\vph$, the electric potential measured with respect to a reference lithium electrode, rather than $\phi$, the true electric potential. In electrochemical applications the potential in an electrolyte is typically measured by inserting a reference electrode of a pure compound. The potential measured depends on the composition of the reference electrode through its chemical potential. Since in lithium battery applications the reference electrode used is nearly always made of lithium, and since much of the data used to calibrate battery models is collected using a lithium reference electrode, it makes sense to use the potential measured with respect to a lithium electrode. Note that $\phi$, the true electric potential, is not a readily measured quantity. In order to switch between $\vph$ and $\phi$ we recall that there is a reversible reaction that occurs on the surface of the electrode between intercalated lithium in the electrode and lithium ions in the electrolyte, given by
\be
\mbox{Li}^+_{(l)}  + \mbox{e}^-_{(s)} \leftrightharpoons  \mbox{Li}_{(s)}. \label{reaction}
\ee
Here the subscript $(l)$ denotes a reactant within the electrolyte and $(s)$ one within the electrode. Typically the current flow into a reference electrode can be assumed to be sufficiently small that this reaction is in quasi-equilibrium. It follows that the electrochemical potentials of compounds on both sides of this equation are equal (\ie $\mu_{p}+ \mu_{\mbox{e}^-}=\mu_{\mbox{Li}}$). Since neither the concentration of electrons nor the lithium within the electrode change,
this equality implies
\bes
\mu_p^0+ RT \log \left( \frac{c}{c_T} \right) +F \phi + \mu^0_{\mbox{e}^-} - F \vph={\rm constant}_1
\ees
(with $+\mu^0_{\mbox{e}^-}$ being the Fermi level of the lithium electrode),  which rearranges to
\be
\phi=\vph- \frac{RT}{F} \log \left( \frac{c}{c_T} \right) +{\rm constant}_2. 
\label{lipot}
\ee
Using (\ref{lipot}) to substitute 
for $\phi$ in (\ref{cn4}) yields the electrolyte Ohm's law found in the Newman theory
\be
\jj=-\hat{\kappa}(c) \left[\nabla \vph -2  \frac{RT}{F} (1- {t_+}) \frac{\nabla c}{c} \right]. 
\label{dil-ohm}
\ee
The effect of the difference in the two different potentials is now readily seen by comparing (\ref{dil-ohm}) with (\ref{cn4}).
Therefore (\ref{dil-ohm}), together with (\ref{cn3}) and (\ref{cn4-div}) is a system of equations equivalent to the charge-neutral equations (\ref{cn1}) and (\ref{cn2}).


\subsection{Moderately-concentrated electrolytes  \label{2} \label{conc}}
Motivated by the confusion in the literature highlighted in \cite{ramos16}, this section reviews, in detail, a commonly used model for moderately-concentrated electrolytes, presented in \cite{newman-book}, which is applicable to most electrolytes consisting of a salt dissolved in a solvent but not to ionic liquids. 
In most practical battery systems ion transport takes place through electrolytic solutions which do not behave as ideal dilute materials as demonstrated primarily by the concentration dependence of their conductivity, see \cite{valoen05}, but also by activity coefficient measurements, e.g. those in \cite{samson99}.
In order to capture this non-ideal behaviour it is necessary to consider not only ion/solvent interactions (as is done in the PNP theory of ideal electrolytes as covered in \S \ref{dil}) but also interactions between the ionic species. Inter-ionic interactions are significant, in even relatively dilute solutions, because the local attraction between oppositely charged ions result in a propensity for ions of opposite charge to lie close to each other, which reduces their mobility in an electric field. This, in turn, reduces the ionic conductivity, see \cite{samson99}. Models of batteries that use electrolytes in this moderately-concentrated regime have been pioneered, and applied successfully to a variety of systems. For example see the series of seminal works by John Newman and his co-workers: \cite{1973Ne,newman-book,doyle93,newman03,doyle96,srinivasan04}.

The electrolyte theory reviewed below is based on the Stefan-Maxwell equations (see, for example, \cite{bird02}), which describe transport in a mixture (including diffusion) in terms of the drag coefficients between its various components.

\subsubsection{Stefan-Maxwell equations \label{smax}}
We start by briefly considering the general case in which the electrolyte is comprised of
$N$ (ionic and solvent) species. \cite{newman-book} uses the Stefan-Maxwell equations as the
foundation of concentrated electrolyte theory. These relate the drag force acting on a component in a mixture to its relative velocity with the other components. It is by balancing these drag forces with gradients of electrochemical potential
of a species and gradients in the fluid pressure that we obtain the average velocity of each species and in turn its flux. {This combined with statements of conservation of species form the equations of moderately concentrated electrolyte theory.}
We note that other mechanisms in addition to the interspecies drag, electrochemical potential and pressure forces can also cause mass transfer and we briefly mention these without detailed dicussion.

In order to derive the force  acting on each component of the mixture as a consequence of the pressure gradient it is neccessary first to obtain an equation of state that relates the concentrations of the $N$ species forming the mixture. According to \cite{liu14}  for most electrolytes it is usually a good approximation to assume that each species has constant molar volume. This is equivalent to the assumption that  the volume occupied by one mole a given species remains fixed whatever the composition of the mixture. On denoting the molar volume of the $i$'th species by $H_i$ we obtain the following equation of state relating the molar concentrations 
\be
\sum_{i=1}^N H_i c_i=1. \label{incompress}
\ee
As we shall see this relation allows us to write down the force on a species arising from the pressure gradient in the mixture. We note that electrolytes may have mechanical properties, such as acting as a viscous fluid or an elastic solid, which can create additional forces. We do not consider these but they can contriubute significantly in certain situations.

The mutual friction force between species
$i$ and $j$ is assumed to be proportional to the friction forces
arising from velocity differences between the species. {Furthermore this force is
proportional to the mole fraction, $\chi_k$, of each
species (see \cite{bothe11}) as defined by 
\be 
\chi_k=\frac{c_k}{c_T} \quad\hbox{ for } k=1...N, \quad \mbox{where} \quad
c_T=\sum_{k=1}^N c_k. 
\label{molefrac} 
\ee 
Here $c_k$ is the molar
concentrations of species $k$ and $c_T$ is the total molar
concentration of all species in the solution. The Stefan-Maxwell
equations give a relation between $\dha_i$, the drag force exerted
on  species $i$, per unit volume of mixture, {and the velocities of
the various species}. In light of the above comments the drag force on
the $i$'th species (per mole unit volume) is taken to depend linearly on the velocity differences between {species}, and modelled (see \cite{bothe11})  by the expression 
\be
\dha_i=RT c_i \sum_{\stackrel{j=1}{j \neq i}}^N k_{ij}
\chi_j (\vect{v}_j-\vect{v}_i)=RT c_T \sum_{\stackrel{j=1}{j \neq i}}^N k_{ij} \chi_i
\chi_j (\vect{v}_j-\vect{v}_i),
\label{eq:MS} 
\ee
 where $\vect{v}_k$
is the velocity of species $k$ and $RT k_{ij}$
is the drag coefficient on one mole of species $i$ moving through pure
species $j$.  Note that $k_{ij}$ is symmetric (\ie $k_{ij}=k_{ji}$) because the drag exerted on species $i$ by species $j$ is equal and opposite to that exerted on species $j$ by species$i$. Here the Maxwell-Stefan inter-species diffusivity
is related to $k_{ij}$ by the Einstein relation so that
$D_{ij}=1/k_{ij}$. }

{The drag force $\dha_i$, is balanced by motive forces (per unit volume) down gradients in the
electrochemical potential $\mu_i$ and down gradients in the pressure $p$
\be
\dha_i-c_i \nabla\mu_i - H_i c_i \nabla p =\vect{0},  \quad\hbox{ for } k=1...N.
\label{eq:MS1} 
\ee
Here the electrochemical potentials $\mu_i$ may be rewritten in terms of the chemical potentials $\mub_i$ and the electric potential $\phi$ in the standard fashion
\be
\mu_i=\mub_i + z_i F \phi,  \quad\hbox{ for } k=1...N, \label{echempot}
\ee
where $z_i$ is the valence of species $i$. The force balance \eqref{eq:MS1} equations are 
supplemented by the standard conservation equations, which can be
  written as 
\be 
\frac{\dd c_i}{\dd t} + \nabla \cdot ( \vect{v}_i
c_i)=0 \qquad \mbox{for} \quad i=1,\cdots,N . 
\label{cons} 
\ee}

{A  force balance on the entire mixture can be obtained by adding together the $N$ relations \eqref{eq:MS1} and noting that  $\sum_{i=1}^N \dha_i=\vect{0}$, it is
\bes
\sum_{i=1}^N c_i \nabla\mu_i + \left(\sum_{i=1}^N H_i c_i\right) \nabla p = \vect{0}.
\ees
By substituting for $\sum_{i=1}^N H_i c_i$ from the equation of state \eqref{incompress} and for the electrochemical potential from \eqref{echempot} we obtain the following expression for the total force balance
\be
\left(\sum_{i=1}^N c_i \nabla \mub_i \right) + \left(\sum_{i=1}^N   F z_i c_i \right) \nabla \phi +\nabla p = \vect{0}. \label{peq1}
\ee
It is straightforward to show from the Gibbs-Duhem relation between the chemical potentials, namely
\bes
\sum_{i=1}^N \chi_i \dd \mub_i/\dd \chi_p=0,
\ees
as derived in Appendix \ref{appa}, that 
\be
\sum_{i=1}^N c_i \nabla \mub_i = \vect{0}.
\ee
This result implies that gradients in the chemical potential, in isolation, do not, as might be expected, lead to a net force on the mixture. In turn this means that the pressure equation \eqref{peq1} can be simplified to
\be
\nabla p = -\rho \nabla \phi  \quad \mbox{where} \quad \rho=\sum_{i=1}^N   F z_i c_i . \label{peq2}
\ee
in which $\rho$ represents the charge density of the mixture. 
}

{It is worth making some brief comments about \eqref{peq2} which gives an intuitively appealing balance between electrostatic forces and pressure forces acting on the mixture. By taking the curl of \eqref{peq2}, it is clear that it can only be satisfied if $\nabla \rho \times \nabla \phi =\vect{0}$ (\ie the gradient of the charge density lies parallel to the electric field $\vect{E}=-\nabla \phi$). There are two special cases where this relation is automatically satisfied which are particularly relevant here. The first of these is where $\rho =0$ (or is negligible), and in this instance no pressure gradient is required to balance the electric force on the mixture and so results in a spatially uniform pressure; this is the case that applies to {\it charge neutral} elecrolytes and so is pertinent to battery modelling. The second special case is where the problem is strictly one-dimensional when the pressure gradient simply counteracts the electrical force on the mixture. This is the same situation as occurs in one dimensional fluid flow where incompressibility dictates that the motion is determined by the
concentrations without reference to any forces and is discussed in the context of other such multiphase systems in \cite{drew82}. 

In more general cases the force balances, described above in \eqref{eq:MS1}, are too naive and need
to be supplemented by multiphase viscous dissipation terms as described in \cite{drew82}.
Such issues, however, are beyond the scope of this work but they are addressed in this context in \cite{liu14}.

A final comment about the general moderately-concentrated problem is that the electrochemical potential of the $i$'th species $\mu_i$ has essentially the same form as that written down for a dilute 1:1 solute in (\ref{ecpots}). The only modification is that the species mole fraction $c_i/c_T$ is replaced by its activity $a_i$ to reflect the fact that the solution is non-ideal. Hence we find
\be
\mu_i=\mu_i^0 + RT \log(a_i) + z_i F \phi ,
\label{EP}
\ee
where $z_i$ is again the charge state of the $i$'th species.

\subsubsection{Stefan Maxwell equations for a binary 1:1 electrolyte}

We now take the general theory of \S\ref{smax} and restrict attention to the case of a 1:1 electrolyte
comprised of a solution of Li$^+$ ions and a generic negative counter
ion species dissolved in a single solvent species. Although
battery electrolytes are often based on rather complex solvent
mixtures, which are usually closely guarded industrial secrets, this
approach provides a reasonable description of many battery
electrolytes. {In Figure \ref{fitting}, we parameterize the model
against experimental data for the most common lithium ion electrolyte
LiPF$_6$ in 1:1 EC:DMC from \cite{valoen05}, treating the two component
solvent (EC:DMC) as if it they were a single solvent.}

In line with \cite{newman-book} we denote the three species making up the electrolyte, namely the solvent, the ${\rm Li}^+$ ions and the generic counterions, by the subscripts $\textit{i}=0,p,n$, respectively. We have therefore $z_0=0$, $z_p=1$ and $z_n=-1$. Combining (\ref{molefrac})-(\ref{eq:MS1}) and expanding in component form yields
\be
-\p \nabla \mu_p={\cal K}_{pn}(\vect{v}_p-\vect{v}_n)+{\cal K}_{p0}(\vect{v}_p-\vect{v}_0) 
+H_p \p \nabla p
\label{eq:p}\\
-\n \nabla \mu_n={\cal K}_{np}(\vect{v}_n-\vect{v}_p)+{\cal K}_{n0}(\vect{v}_n-\vect{v}_0) 
+H_n \n \nabla p
\label{eq:n}\\
-\s \nabla \mu_0={\cal K}_{0p}(\vect{v}_0-\vect{v}_p)+{\cal K}_{0n}(\vect{v}_0-\vect{v}_n) 
+H_0 \s \nabla p
\label{eq:s}
\ee
where, by using (\ref{molefrac}) (\ie $c_T=\s+c_p+c_n$) and (\ref{eq:MS})-(\ref{eq:MS1}), the drag coefficients can be expressed in the form
\be
{\cal K}_{ij}=RT\frac{c_i c_j}{c_T D_{ij}} = {\cal K}_{ji}, \qquad \mbox{with} \quad i=0,p,n,\quad j=0,p,n.
\label{eq:K}
\ee
Henceforth we can omit (\ref{eq:s}), noting that the choice of physically realistic functions $\mu_k$ (with $k=0,n,p$) ensures this is satisfied.
Using (\ref{EP}) the electrochemical potentials of the ion species are
\be
 \mu_{n}=\mu_n^0 + RT \log(a_n) - F\phi, \qquad
 \mu_p=\mu_p^0 + RT \log(a_p) + F\phi. 
\label{eq:pot}
\ee
As in dilute theory, charge neutrality can be assumed so that we can write
\be
\n= \p= c .\label{electroneut}
\ee
As discussed above, where the electrolyte is charge neutral (\ie $\n=\p$), the solution to the pressure equation \eqref{peq2} is such that $p$ is constant and hence the pressure gradient terms in \eqref{eq:p}-\eqref{eq:s} vanish.

We now seek to find a constitutive equation for the current density $\jj$. We will broadly follow the derivation given in \cite{newman-book} but attempt to clarify their argument. As in the dilute case, the current density is given by (\ref{curr}), which can be rewritten as
\be
\jj=Fc(\vect{v}_p-\vect{v}_n) .
\label{eq:newv}
\ee
Substitution of (\ref{eq:newv}) into (\ref{eq:p}) and (\ref{eq:n}), {taking account of (\ref{electroneut}) and the fact that the pressure gradient terms vanish}, yields
\be
-c\nabla \mu_p={\cal K}_{p0}(\vect{v}_p-\vect{v}_0)+\frac{{\cal K}_{pn}}{Fc} \jj \label{eq:cv1}\\
-c\nabla \mu_n={\cal K}_{n0}(\vect{v}_n-\vect{v}_0)-\frac{{\cal K}_{np}}{Fc} \jj \label{eq:cv2},
\ee
where
\be
{\cal K}_{pn}={\cal K}_{np}=\frac{RT c^2}{c_T D_{pn}}, \quad {\cal K}_{p0}=\frac{RT c \s}{c_T D_{p0}}, \quad {\cal K}_{n0}=\frac{RT c \s}{c_T D_{n0}} \ \  \mbox{and} \ \ c_T=(\s+2c).
\ee
Equations (\ref{eq:newv}), (\ref{eq:cv1}) and (\ref{eq:cv2}) can be re-arranged to give expressions for the ion velocities in terms of the solvent velocity
\be
\vect{v}_p=\vect{v}_0-\frac{c_T}{RT}\frac{D_{p0}}{\s}\nabla \mu_p-\frac{D_{p0}}{D_{pn}F \s}\jj \label{eq:v1},\\
\vect{v}_n=\vect{v}_0-\frac{c_T}{RT}\frac{D_{n0}}{\s}\nabla \mu_n+\frac{D_{n0}}{D_{pn}F \s}\jj \label{eq:v2}.
\ee
Subtracting (\ref{eq:v2}) from (\ref{eq:v1}) gives
\be
\vect{v}_p - \vect{v}_n=\frac{c_T}{RT \s}(D_{n0}\nabla\mu_n-D_{p0}\nabla\mu_p)-\frac{D_{p0}+D_{n0}}{F \s D_{pn}}\jj. \label{eq:j2}
\ee
On substituting for $\vect{v}_p-\vect{v}_n$ (in terms of $\jj$) from (\ref{eq:newv}), and for $\mu_p$ and $\mu_n$ from (\ref{eq:pot}), and rearranging the resulting expression we obtain the following expression for $\jj$:
\be
\jj =  -\kappa(c) \left( \nabla \phi + \frac{RT}{F} ( \tn \nabla \log(a_p) - (1-\tn) \nabla \log(a_n)) \right),
\label{eq:nablamu2}
\ee 
where 
\be 
\tn=\frac{D_{p0}}{D_{p0}+D_{n0}}, \qquad \kappa(c)=
\frac{F^2 c_T D_{pn} c (D_{p0}+D_{n0})}{RT(c(D_{p0}+D_{n0})+D_{pn} \s )} 
\label{eq:trans} 
\ee 
are the transference number of the (positive) lithium ions with respect to the solvent velocity, and the conductivity of the electrolyte as a function of the concentration. {Note how this equation for $\jj$, and the definition of transference number and conductivity compare to the dilute version given in (\ref{cn4})-(\ref{cn4b}).

We now rewrite {equations \eqref{eq:v1}--\eqref{eq:v2}} in a form that avoids the use of a chemical potential for each ion species and instead uses a chemical potential for the entire electrolyte $\mu_e$.
Without any loss of generality we can express $\vp$ and $\vn$ in the form
\be
\begin{array}{l}
\ds \vp=((1-\alpha) \vp+\alpha \vn) +\alpha(\vp-\vn), \\*[2mm]
\ds \vn=((1-\alpha) \vp+\alpha \vn)-(1-\alpha)(\vp-\vn),
\end{array}
\label{alpha}
\ee
for any function $\alpha$. Our procedure is to substitute $\vp-\vn=\jj/(Fc)$ in the final terms of these expressions and then to replace $\vn$ and $\vp$, everywhere else on the right-hand side of these expressions, from (\ref{eq:v1}) and (\ref{eq:v2}). Choosing $\alpha=\tn$, as defined in (\ref{eq:trans}), allows the electric potential $\phi$ to be eliminated from the term $(1-\alpha) \vp+\alpha \vn$. 
The expressions for $\vp$ and $\vn$, in (\ref{alpha}),  can then be rewritten in the form
\be
\vect{v}_p=\vect{v}_0-\frac{c_T}{RT \s}\frac{D_{n0}D_{p0}}{D_{p0}+D_{n0}}(\nabla \mu_p+\nabla \mu_n)+\frac{\tn}{Fc}\jj , 
\label{lobster1}\\
\vect{v}_n=\vect{v}_0-\frac{c_T}{RT \s}\frac{D_{n0} D_{p0}}{D_{p0}+D_{n0}}(\nabla \mu_p+\nabla \mu_n)-\frac{(1-\tn)}{Fc}\jj .
\label{lobster2}
\ee
These equations can be further simplified by introducing the electrolyte chemical potential $\mu_e(c)$ (a function of electrolyte concentration only) and  the chemical diffusion coefficient $\dc$, as we did in (\ref{cn3}), defined by
\be
\mu_e=\frac{\mu^0_n+\mu^0_p}{2}+RT \log((a_n a_p)^{1/2}) \quad \mbox{and} \quad \dc=\frac{2D_{n0}D_{p0}}{D_{n0}+D_{p0}}, \label{eq:diffmu}
\ee
and noting that, from (\ref{eq:pot}), $\mu_e=\frac{1}{2}(\mu_p+\mu_n)$. We then find that (\ref{lobster1})-(\ref{lobster2}) can be rewritten in the form
\be
\vect{v}_p &=&\vect{v}_0-\frac{c_T}{\s RT} \dc\nabla \mu_e+\frac{\tn}{Fc}\jj, \label{eq:v11}\\
\vect{v}_n &=&\vect{v}_0-\frac{c_T}{\s RT} \dc\nabla \mu_e-\frac{(1-\tn)}{Fc}\jj.
\label{eq:v22}
\ee
Notably, $D_{n0}$ and $D_{p0}$ can vary with concentration independently, 
without affecting the preceding analysis. It follows that $\dc$ and $\tn$ may also vary independently as functions of concentration.

The remaining equations governing the behaviour come from considering conservation of the ions and solvent as well as the volume they occupy.
The mass conservation equations for the two ion species are the same as (\ref{cons}), namely
\be
\frac{\dd c}{\dd t}+\nabla\cdot(c\vect{v}_p)=0, \quad  
\label{eq:conserve1}
\frac{\dd c}{\dd t}+\nabla\cdot(c\vect{v}_n)=0. 
\label{eq:conserve2}
\ee
Taking the differences of these two equations, and substituting for $\vect{v}_p-\vect{v}_n$ from (\ref{eq:newv}), yields an equation for current conservation
\be
\nabla \cdot\jj=0. 
\label{eq:chargcons}
\ee
Substituting  $\vect{v}_p$ (from (\ref{eq:v11})) in (\ref{eq:conserve1}) yields\footnote{The same result is obtained by substituting for $\vn$ (from (\ref{eq:v22})) in (\ref{eq:conserve2}).}
\be
\frac{\dd c}{\dd t} - \nabla\cdot\left( \frac{c_T}{\s RT}c{\cal D} \nabla \mu_e\right)+\nabla\cdot(c\vect{v}_0)=-\frac{\nabla\tn\cdot \jj}{F}  
\label{eq:diffconserve}
\ee 
and this can be compared to its dilute theory counterpart given in
(\ref{cn3})). {These equations couple to the} mass conservation equation for
the solvent which, from (\ref{cons}), is given by 
\be 
\frac{\dd \s
}{\dd t}+\nabla\cdot(\s \vect{v}_0)=0. \label{eq:conserve3} 
\ee 
{Finally we
require an equation of state. On using the charge neutrality condition $\n=\p=c$  in \eqref{incompress} we find that this can be written in the form
\be 
H_c c+H_0 \s = 1, 
\label{state} 
\ee 
where $H_c=H_n+H_p$.}

In one dimension we now have sufficient equations to
specify the problem; these are composed of the five equations (\ref{eq:nablamu2}),
(\ref{eq:chargcons}), (\ref{eq:diffconserve}), (\ref{eq:conserve3})
and (\ref{state}) for the five variables $c$, $\s$, $\phi$, $\jj$ and
$\vect{v}_0$. {Note that, in more than one dimension, these equations are not sufficient, at least not in general, as can be seen by multiplying \eqref{eq:conserve1} by $H_c$ and adding to \eqref{eq:conserve1} multiplied by $H_0$. On using \eqref{state} to eliminate the time-derivative from the resulting equation this yields the scalar PDE
\be 
\nabla \cdot ( c H_c \vect{v}_p + c_0 H_0 \vect{v}_0 )=0,
\ee 
which in multiple dimensions is insufficient to determine the vector $\vect{v}_0$. As has been described in \cite{drew82} this conservation equation needs to be supplemented by a momentum equation, but this is beyond the scope of this work.
}

Having posed the governing equations\textcolor{red}{,} an additional simplifying assumption is commonly made, which appears to be adequate for most solvent based battery electrolytes. {This consists of assuming that the electrolyte is sufficiently dilute so that $\s \approx c_T$ in which case \eqref{state} can be approximated by $\s \approx 1/H_0$ and the solvent velocity is small (\ie $|\vect{v}_0| \ll |\vect{v}_p|$ and $|\vect{v}_0| \ll |\vect{v}_n|$). This limit is equivalent to the approximation}
\be
\vect{v}_0 \equiv \vect{0}  \label{vw0}
\ee
and only requires
solution of the three equations (\ref{eq:nablamu2}), (\ref{eq:chargcons}) and (\ref{eq:diffconserve}) for the three variables $c$, $\phi$ and $\jj$ {(instead of five equations for the five variables in the full model)}. {This assumption also avoids the complication, that arises in multiple dimensions, of requiring to be supplemented by a momentum equation. Note however that, even in this small concentration limit, we still include interphase drag between negative and positive ions. This is because interphase drag is very often significant even at quite at low concentrations (often as low as 0.1 molar which is usually a small mole fraction). It arises because positive and negative ions interact via a strong long-range force (the Coulomb force).}

\subsubsection{The potential measured with respect to a lithium electrode.}
As in the dilute case it is useful to reformulate the model in terms of $\vph$, the potential measured with respect to a lithium electrode.  Once again we assume that the reaction (\ref{reaction}) occurring on the surface of the reference electrode is in quasi-equilibrium, so that sum of the electrochemical potentials of compounds on each side of (\ref{reaction})  are identical (\ie $\mu_{p}+ \mu_{e^-}=\mu_{\text{Li}}$). However, owing to the slightly different definition  of $\mu_p$ in the moderately-concentrated case (compare (\ref{eq:pot}) with (\ref{ecpots})),  this leads to a modified version of the relation between $\phi$ and $\vph$, namely
\be
\phi=\vph-\frac{RT}{F} \log(a_p) + {\rm constant},
\ee
which can be compared to (\ref{lipot}). On substitution of this expression for $\phi$ into (\ref{eq:nablamu2}) we can re-express the constitutive equation for the current density equation in the form
\bes
\jj =-\kappa(c)\left(\nabla \vph - \frac{RT}{F}(1-\tn)(\nabla \log(a_p)+\nabla\log(a_n))\right),
\ees
which, on referring to the definition of the electrolyte chemical potential in (\ref{eq:diffmu}), can be re-expressed as
\be
\jj =-\kappa(c) \left(\nabla \vph-\frac{2}{F}(1-\tn)\nabla \mu_e\right).   \label{ohm}
\ee
This is a clearer way to write Ohm's law than (\ref{eq:nablamu2}) because it allows $\jj$ to expressed solely in terms of the measurable quantities $\vph$ and $\mu_e$ (notably the activities of the individual ion species $a_n$ and $a_p$ are not directly measurable). This expression differs slightly from that typically presented by Newman and co-authors because they tend to use the dilute limit of (\ref{ohm}), namely (\ref{dil-ohm}).


\subsection{Summary: model for a moderately-concentrated electrolyte }
The moderately-concentrated theory is relevant in those situations where  the salt concentration is small compared to the solvent concentration so that $c{\ll}c_T$, the interactions forces are large $(D_{n0} + D_{p0}){\gg}D_{pn}$, and the solvent velocity can be neglected $\vect{v}_0 {\approx} 0$. In this limit the moderately-concentrated charge transport model  (\ref{eq:chargcons}), (\ref{eq:diffconserve}) and (\ref{ohm}) takes the form
\be
\frac{\dd c}{\dd t} - \nabla\cdot\left( {c\dc(c) } \nabla \left( \frac{\mu_e(c)}{RT} \right)\right) &=&-\frac{\nabla \tn\cdot \jj}{F} ,
\label{mcon1}\\
\nabla \cdot \jj &=& 0 ,
\label{mcon2}\\
\jj &=&-\kappa(c) \left(\nabla \vph-\frac{2 RT}{F}(1-\tn)\nabla \left( \frac{\mu_e(c)}{RT} \right) \right),  
\label{mcon3}
\ee
where
\be
\dc(c)=\frac{{2}D_{n0} D_{p0}}{D_{n0} + D_{p0}} \quad \mbox{and} \quad \kappa(c)=\frac{F^2 c}{RT} \frac{(D_{n0} + D_{p0})}{\left(1+\frac{D_{n0} + D_{p0}}{D_{pn}} \frac{c}{c_T} \right)}.
\ee
Note that if we consider the limit of a dilute electrolyte with $c{\ll}c_T$,
and the interaction forces are not large
$(D_{n0} + D_{p0})/D_{pn}=O(1)$, then
we revert to the dilute
solution model (\ref{cn3}) and (\ref{dil-ohm}) provided that we
identify $D_{p0}$ with $D_p$ and $D_{n0}$ with $D_n$.

It is worth noting that (\ref{mcon1})-(\ref{mcon3}) can be rewritten in the form
\be
\frac{\dd c}{\dd t} - \nabla\cdot\left( \de(c) \nabla c \right) &=&-\frac{\nabla \tn\cdot \jj}{F} , 
\label{mcon4}\\
\nabla \cdot \jj &=& 0 ,
\label{mcon5}\\
\jj &=&-\kappa(c) \left(\nabla \vph-\frac{2 RT}{F}(1-\tn) \frac{a_e'(c)}{a_e(c)}\nabla c \right),  
\label{mcon6}
\ee
where $a_e(c)=(a_n a_p)^{1/2}$ is the activity coefficient of the electrolyte (such that $\mu_e(c)=\mu_e^0+RT \log( a_e(c))$) and the effective diffusivity is given by
\bes
\de(c)= \dc(c) \frac{c a_e'(c)}{a_e(c)}.
\ees
An implicit assumption in Newman's formulation of the equations is that the salt solution behaves as an ideal solution so that ${a_e'(c)}/{a_e(c)}=1/c$.


To fit the model it is necessary to determine the lithium diffusivity
$\de(c)$, the electrolyte conductivity $\kappa(c)$ and the transference number
$\tn(c)$. Doing this function fitting leads to a relatively robust way
of modelling the electrolyte for engineering applications. An example
of fitting the phenomenological model functions $\de(c)$ and
$\kappa(c)$ to data is shown in Figure \ref{fitting} for the
electrolyte LiPF$_{6}$ in 1:1 EC:DMC at $T=293$K (this data comes from
\cite{valoen05}). For this electrolyte the transference
number {of lithium ions, with respect to the solvent
velocity,} is found to be approximately constant with $\tn=0.38$.

\begin{figure}[ht!]
\begin{center}
\includegraphics[width=0.49\columnwidth]{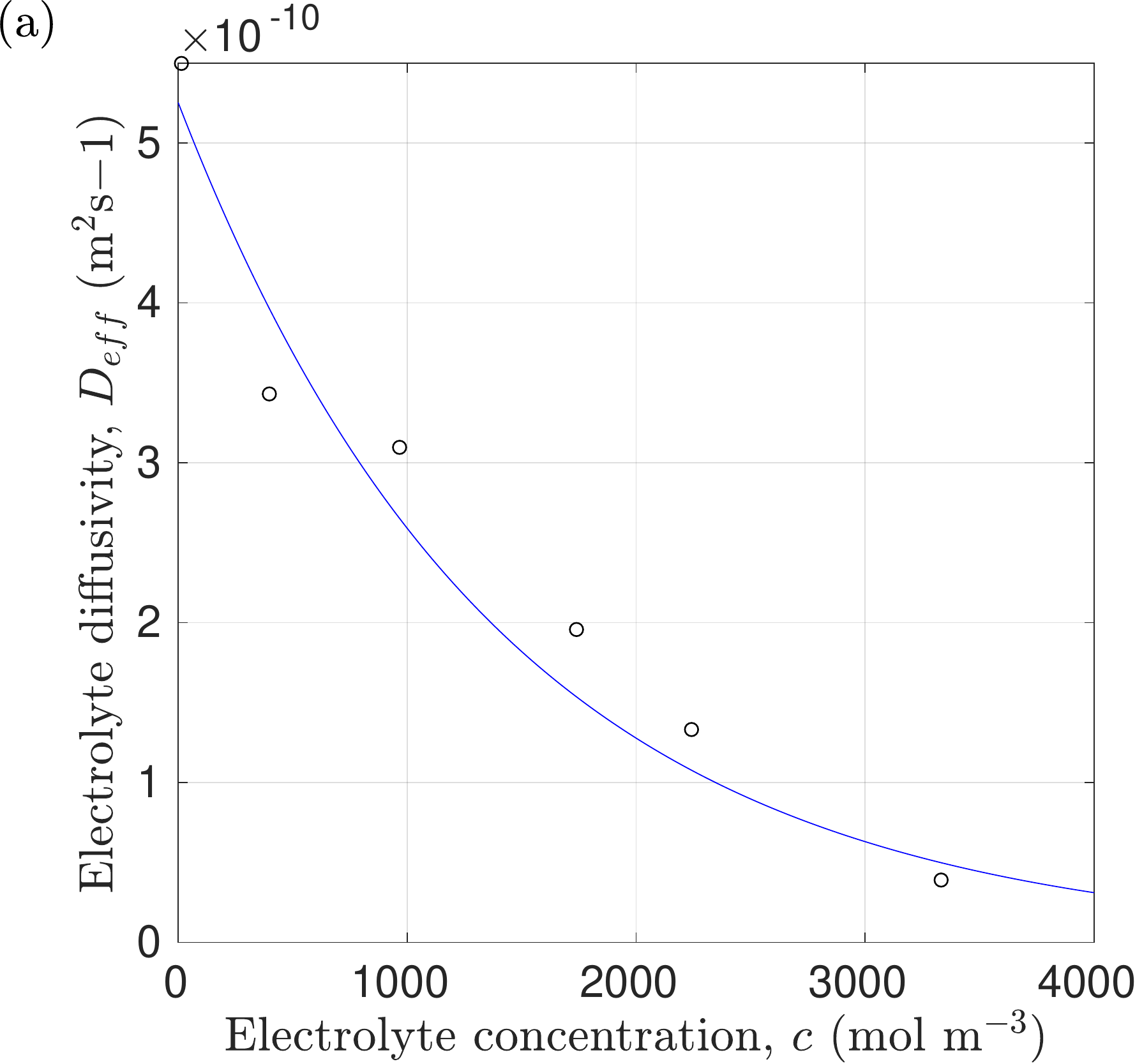}\includegraphics[width=0.49\columnwidth]{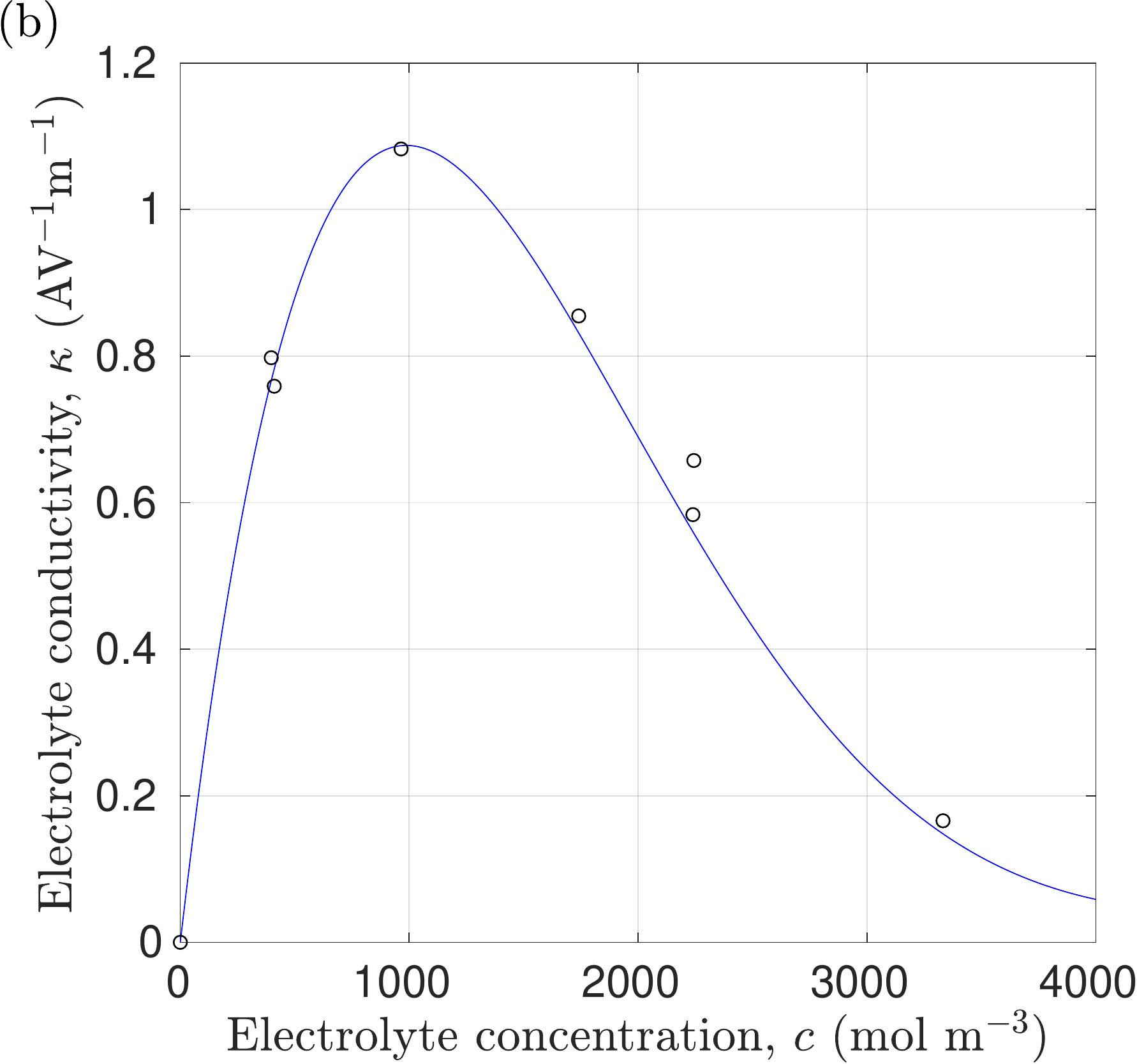}
\caption{\label{fitting} (a) Diffusion coefficient $\de$ for LiPF$_{6}$ in 1:1 EC:DMC at $T=293$K as a function of concentration and (b) electrolyte conductivity $\kappa$ for the same electrolyte. Lines represent the fit to to the experimental data (circles) taken from \cite{valoen05}. The fitted function for the diffusivity and conductivity are given by $\de(c)=5.3\times 10^{-10} \exp(-7.1\times 10^{-4}c)$ and $\kappa(c)=10^{-4} c (5.2-0.002c+2.3\times10^{-7}c^2)^2$ respectively.}
\end{center}
\end{figure}


\section{Lithium transport and electric current flow through the electrode particles \label{particles}}

Here we review the modelling of lithium transport in individual electrode particles and current transport through the solid matrix formed by the agglomeration of electrode particles, polymer binder material and conductivity enhancers (such as carbon black). Transport of lithium through the microscopic electrode particles is typically slow; the diffusion timescale for a lithium ion traversing a microscopic electrode particle frequently being comparable to, or even longer, than that required for a lithium ion to traverse the whole cell in the electrolyte. Since the concentration of lithium ions at the surface of the electrode particles strongly influences the rate at which lithium ions are intercalated into the electrode particles from the electrolyte (or {\it vice-versa}) a battery charge transport model  must treat both microscopic transport of lithium through the electrode particles and its macroscopic transport thorough the electrolyte. The coupling between these micro- and macro-scale transport processes occurs via a reaction rate condition which specifies the rate of lithium intercalation (or de-intercalation) at the particle surface in terms of (a) the lithium concentration $c_s$ on the surface, (b) the potential difference $\phi_s-\phi$ between the particle surface and the adjacent electrolyte\footnote{
The potential changes smoothly through the double layer in the electrolyte immediately adjacent to the particle surface, but since the electrolyte model discussed in \S\ref{electrolyte} does not explicitly treat these extremely thin double layers, a potential difference between electrode particle and electrolyte appears in this model due to the potential drop across the double layer.},
and (c) the lithium concentration $c$ in the adjacent electrolyte. The condition that is typically used in practice is the Butler-Volmer relation (see, for example, \cite{bockris70}), which accounts for both the reaction rates and the effects of the double layer, and is based on the quantum mechanical Marcus Theory described in \cite{marcus65}. Usually the potential in the electrode matrix $\phi_s$ and the corresponding current density flow $\jj_s$ through the matrix is modelled by the macroscopic Ohm's Law
\be
\jj_s=-\kappa_s \nabla \phi_s, \label{ohm-solid}
\ee
where $\kappa_s$ is the effective conductivity of the electrode matrix (formed of electrode particles, binder and conductivity enhancer).


\subsection{The standard approach}
Here we set out the standard approach to modelling lithium transport within the electrode particles of a lithium ion cell and the current transfer process between the electrode particles and the surrounding electrolyte. This approach is widely adopted in the modelling literature (\eg \cite{dargaville10,doyle93,doyle96,fuller94,ma95,newman03,srinivasan04}) although it has recently been challenged by an alternative approach which is discussed in \S \ref{cahn}.

The central idea of the model is to consider a one dimensional problem between the two current collectors describing behaviour of the electrolyte and the charge transport in the solid electrodes. The lithium motion within the particles of the electrodes is on a much smaller scale and movement of this is described using a separate dimension representing position within each particle. Hence we will use $x$ to represent the position between the current collectors and $r$ to represent the position within any particle. The problem is therefore often referred to as a pseudo two-dimensional model. More precisely, one might describe the structure of the model as being multiscale, where both the micro- and macroscopic models are one-dimensional. Here we present the basic model but later we describe how it can be systematically derived. 

\subsubsection{Microscopic lithium transport models in individual electrode particles \label{litrans}}
At its very simplest, transport of intercalated lithium within electrode particles is modelled by linear diffusion in an array of uniformly sized (radius $a$) spheres (see \eg \cite{doyle96}). Hence the lithium concentration in an electrode particle at position $x$ within the electrode, $c_s(r,x,t)$, evolves according to
\be
\frac{\dd c_s}{\dd t}= \frac{1}{r^2} \frac{\dd}{\dd r} \left( D_s r^2 \frac{\dd c_s}{\dd r} \right), \qquad  \label{lindiff}
\ee
where $r$ measures distance from the particle's centre and $D_s$ is the solid phase Li diffusion coefficient. This model for lithium transport in the electrode particles is typically coupled to the processes taking place in the adjacent electrolyte through a Butler-Volmer relation, which gives the transfer current density $j_{\rm tr}$ flowing out through the surface of the particle, in terms of the lithium concentrations on the particle's surface $c_s$, the adjacent electrolyte concentration $c$ and the potential difference between the electrolyte and the electrode particle $\vph-\phi_s$. Following Faraday's laws of electrolysis, the flux of lithium on the particle surface is proportional to the current density, leading to the following boundary condition on the electrode particle surface
\bes
 -D_s \frac{\dd c_s}{\dd r}= \frac{1}{F}\;j_{\rm tr} \qquad\hbox{ at } r=a.
\ees
Lithium transport within electrode particles is often better described by nonlinear diffusion (with $D_s=D_s(c_s)$), an approach that is used in \cite{karthikeyan08,farkhondeh11,krach18}, for example. Experimental work also demonstrates the strong relationship between diffusivity and lithium ion concentration within certain materials, such as graphite \cite{takami95,verbrugge03,Levi03,Baker12} and {LiNi$_x$Mn$_y$Co$_{1-x-y}$O$_2$, often
referred to as NMC \cite{wu12,EckerI,EckerII},} a positive electrode material discussed in detail in \S3.3.2. We will also discuss in \S\ref{common} and \S\ref{cahn}
more complicated models of lithium transport applicable to cases where phase separation occurs or the behaviour is highly anisotropic. 

\subsubsection{The Butler-Volmer relation \label{bv}}
The transfer current density $j_{\rm tr}$ is the normal component of the current density on the particle surface, from the electrode particle into the surrounding electrolyte, {and can be} expressed in terms of a Butler-Volmer relation of the form
\begin{equation}
j_{\rm tr} ={i_{0}}(c_s,c)\left(\exp\left[\frac{\alpha_{\rm a} F}{RT}(\phi_s - \vph-U_{\rm eq}(c_s,c))\right]-\exp\left[\frac{-\alpha_{\rm c} F}{RT}(\phi_s - \vph-U_{\rm eq}(c_s,c))\right]\right),
\label{DBV}
\end{equation}
{with $U_{\rm eq}$ depending usually only on $c_s$}. Here $i_0(c_s,c)$ is the exchange current density and $U_{\rm eq}(c_s,c)$ is
the open circuit potential (OCP). Both $i_o$ and $U_{\rm eq}$ depend on the
electrode material while the dimensionless constants $\alpha_{\rm a}$ and
$\alpha_{\rm c}$ are anodic and cathodic transfer coefficients lying between 0 and 1, and are conventionally both taken to be 1/2. 
Note that when measurements of the OCP of a material are taken, they are done so with the use of a reference electrode which is typically lithium. Hence, the potential difference across the double layer is difference between the electric potential $\phi_s$ in the solid and $\vph$ in the electrolyte (the potential measured relative to a lithium electrode). Note this this has been assumed in writing (\ref{DBV}) and as in line with the discussion given below (\ref{lindiff}).

The  OCP is found by considering equilibrium, when $j_{\rm tr}=0$, and
measuring the potential difference between the electrolyte and the
electrode particle, $\phi_s-\vph$, for various different levels of the
concentrations. For most materials there is a maximum concentration of
Li that can occur, denoted by $c_{\rm s,max}$, and the OCP is typically plotted as a function of ${y}=c_s/c_{\rm s,max}$, which is the {Lithium stoichiometry}. The OCP varies widely depending on the electrode material and to
illustrate this in Figure \ref{equilibrium_U}(a) it is plotted for the
lithiated graphite Li$_x$C$_6$ while in Figure
\ref{equilibrium_U}(b) it is plotted for lithiated iron phosphate
Li$_y$FePO$_4$. Note the  OCP for Li$_y$FePO$_4$
has a large almost flat plateau symptomatic of a two-phase state
within the material.

\begin{figure}[ht!]
\begin{center}
\includegraphics[width=0.32\columnwidth]{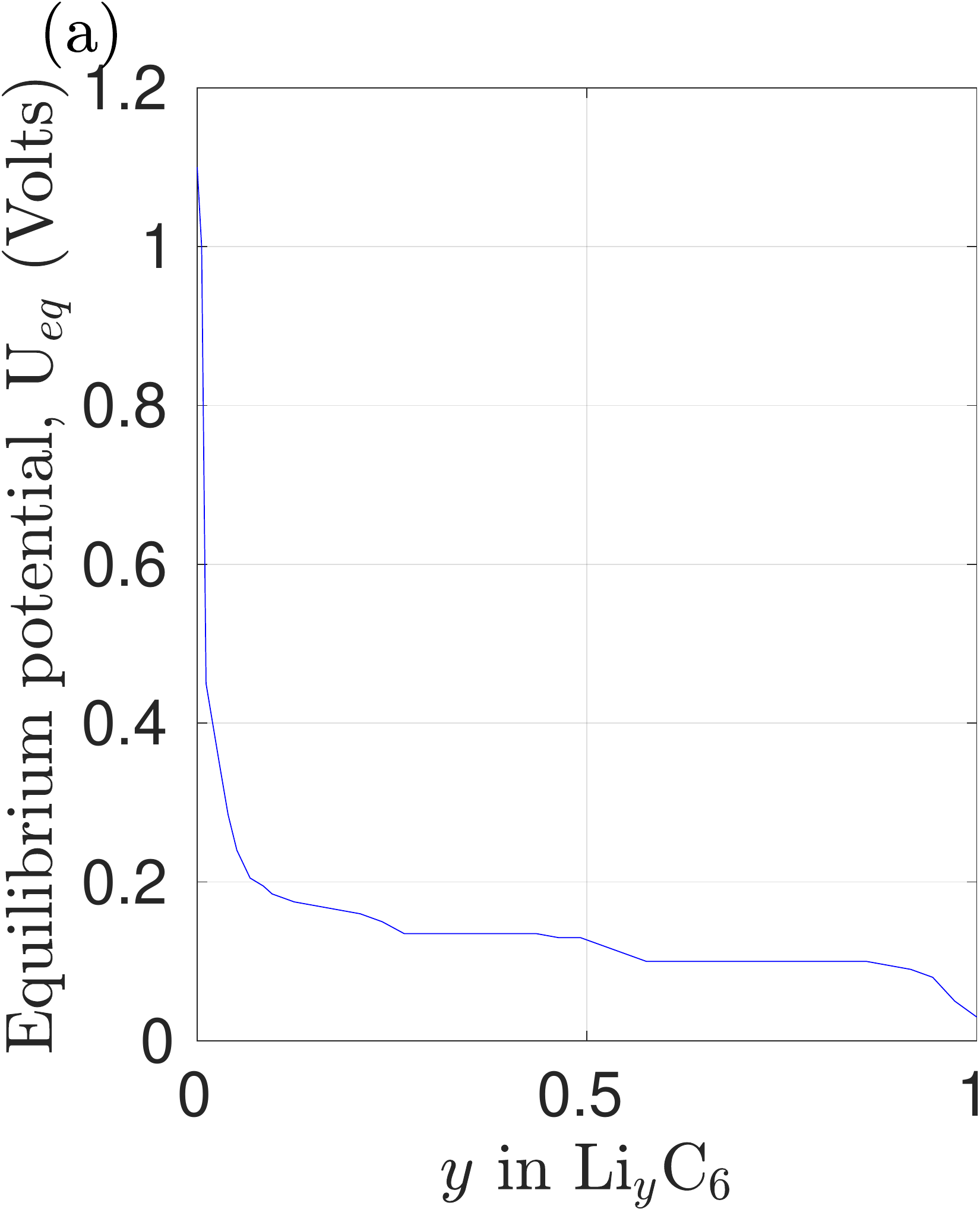}\includegraphics[width=0.32\columnwidth]{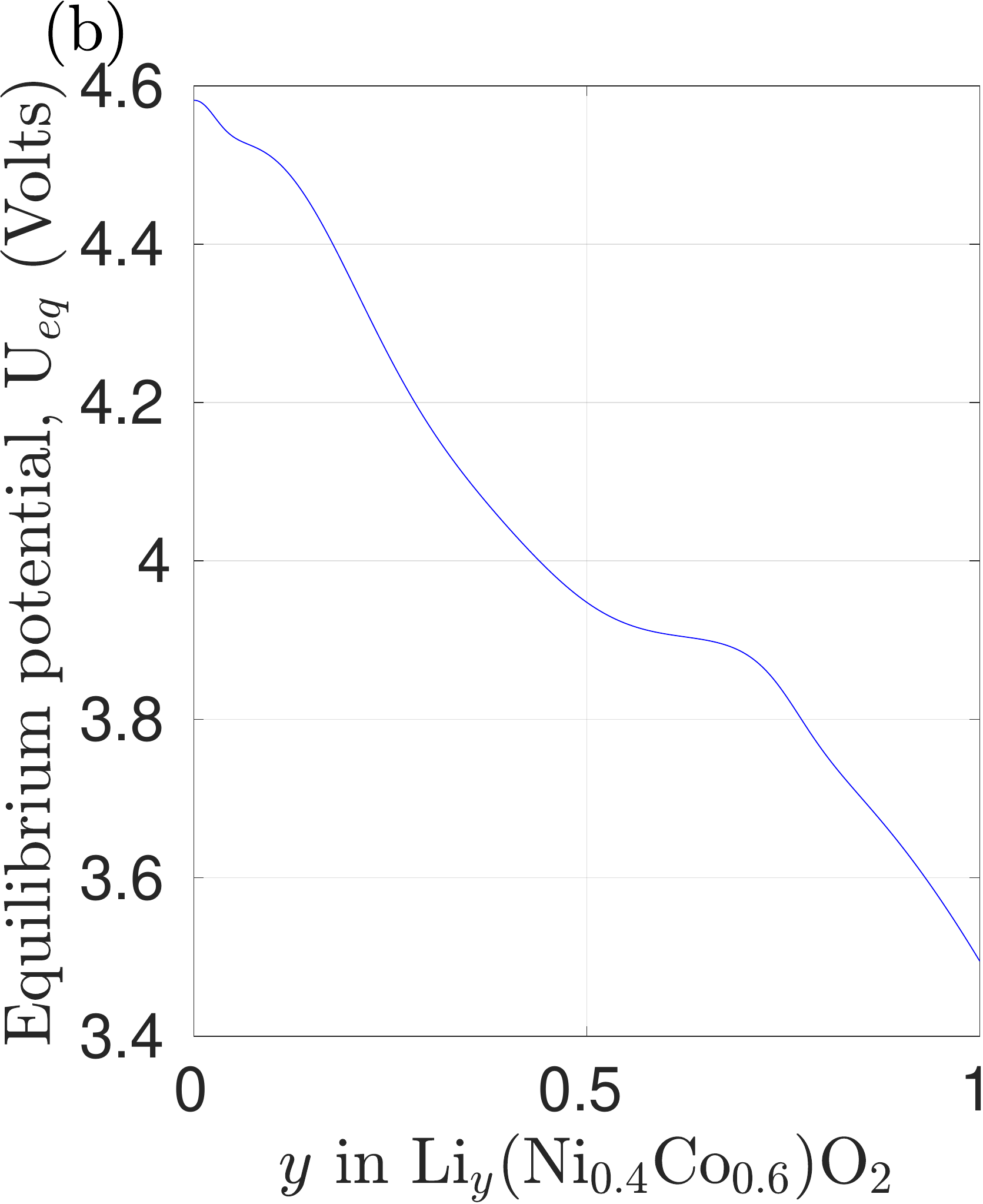}\includegraphics[width=0.32\columnwidth]{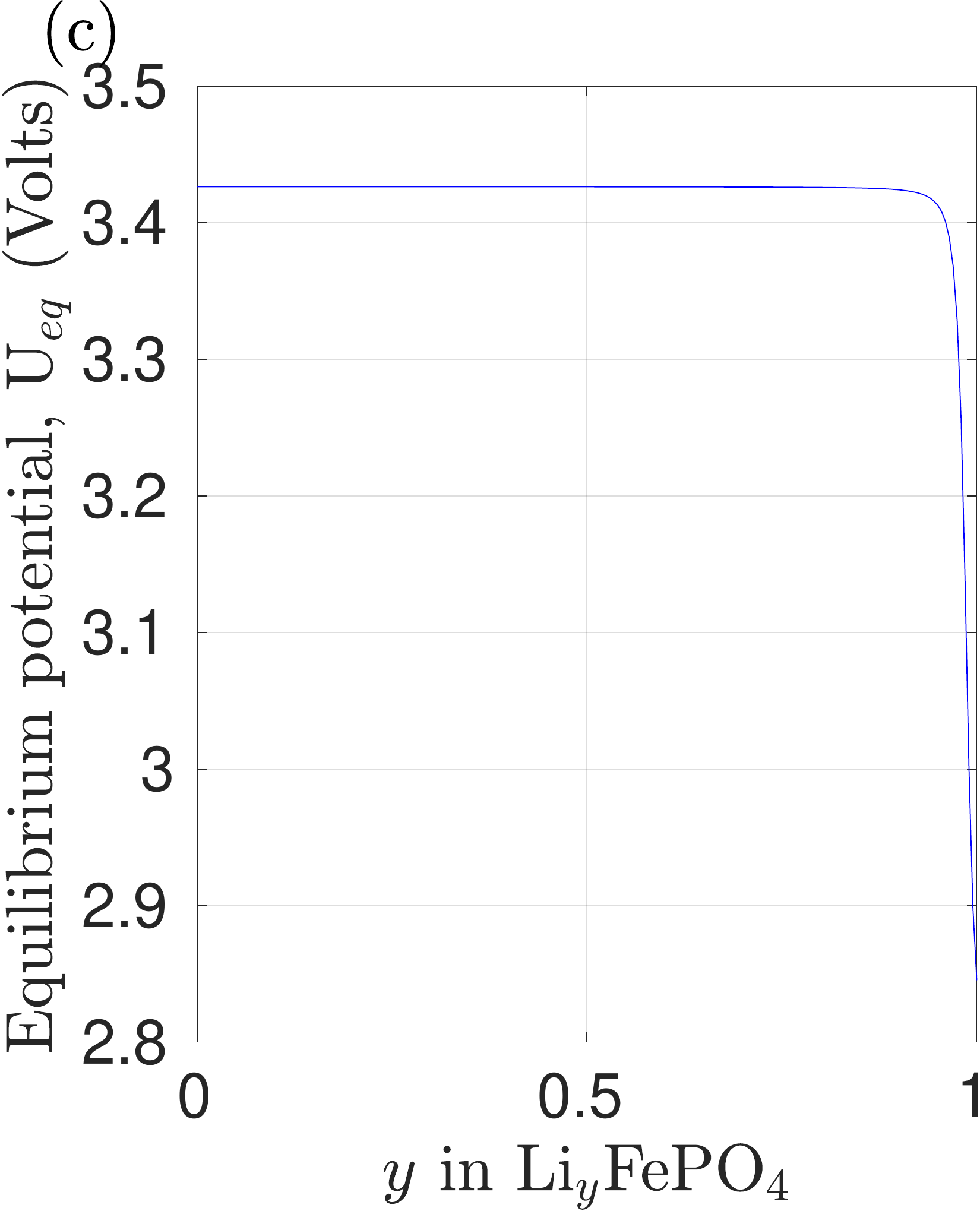}
\caption{\label{equilibrium_U} 
The OCP, $U_{\rm eq}$, of (a) LiC$_6$, from \cite{fuller94}, (b) Li(Ni$_{0.4}$Co$_{0.6}$)O$_2$, from \cite{EckerI}, and (c) LiFePO$_4$, from \cite{srinivasan04}). Each is shown as a function of {Lithium stoichiometry}.
}
\end{center}
\end{figure}

The exchange current $i_0$ is much less well quantified in experiments and commonly is taken to be given by the formula
\begin{equation} 
 i_0 =
  k \;c^{\alpha_{\rm a}}\; (c_{\rm s,max} -c_{\rm s})^{\alpha_{\rm a}} \;c_{\rm s}^{\alpha_{\rm c}},
\label{ddio}
\end{equation}
where $k$ is a kinetic rate constant.


\subsection{Appropriate forms for the Butler-Volmer relation} 
\label{sdbve}
The Butler-Volmer relation is widely used in the literature
(see, for example, \cite{doyle93,1994FuDoNe,2002GoEtAl,2006SmWa,2006-b-SmWa,2007SmRaWa,2010ChEtAl,2012KiSmIrPe}),
and here we focus on why it is necessary to take care in selecting the 
functions used to describe $U_{\rm eq}$ and $i_0$. In the literature
(see, for instance, \cite{2014WeSuPe}) $U_{\rm eq}$ is typically fit
to some polynomial, or other family of functions, such as exponentials. Furthermore, as explained in
\cite{2014WeSuPe}, most of the fitting is done to data points which lie in the 
middle of the stoichiometry range, namely 0.1 -- 0.9 where many batteries operate,
thereby avoiding extreme cases which are prone to triggering failure and/or accelerated degradation, see \cite{wang2012,birkl17}.

Arguably more important than the behaviour for intermediate charge states is the what happens when the electrodes get
close to being fully intercalated $c_{\rm s}=c_{\rm s,max}$, or fully
depleted $c_{\rm s}=0 $, or where the electrolyte approaches depletion
$c=0$. To discuss the allowable behaviours in these cases we consider the scenario in which
$c_{\rm s} \to 0$ and then indicate how the same ideas can be extended to apply to the other limiting cases.

The basic difficulty is that as $c_{\rm s} \to 0$ we need the Li flux
out of the particle to go to zero to prevent predicting nonphysical
negative concentrations in the solid. However, if we start with a
depleted particle and do not allow Li to enter, then it will never
charge.  We need a model that avoids both physically unrealistic
situations.  It is of course always possible to simply chose $i_0$ and
$U_{\rm eq}$ and then impose some switching logic to turn the flux on
or off as required to avoid such problems. However, it is preferable
to have a Butler-Volmer relation that incorporates mechanisms that automatically ensures
such physically unrealistic situations cannot occur.

It is common to consider the two parts of a Butler-Volmer relation in (\ref{DBV}) as representing {reverse and forward} reactions, which are given by the first and second term of the right hand side of that relation, respectively. As
$c_{\rm s} \to 0$, {the reverse (anodic)
reaction should be first order, while the forward (cathodic) should be bounded and
positive (like a zeroth order reaction)}. Such behaviour can be readily achieved
by taking $i_0$ and $U_{\rm eq}$ to have the local form, when
$c_{\rm s} \to 0$,
\begin{equation}
i_0\sim
\left(
  k_{1}\;c_{\rm s}\right)^{\left(\frac{\alpha_{\rm c}}{\alpha_{\rm a}+\alpha_{\rm c}}\right)}
\ \
\mbox{ and }
\ \
 U_{\rm eq}\sim
 \frac{-RT}{F(\alpha_{\rm a}+\alpha_{\rm c})}\log \left( k_{2}\;c_{\rm s} \right),
\end{equation}
where $k_{1}$ and $k_2$ are positive constants (which can be different for each electrode and can also change inside non-homogeneous electrodes). Such a
formulation near this extreme of the concentration will automatically
ensure the flux cannot become positive as the
concentration reduces, thereby avoiding negative concentrations, and
the flux can be finite and positive so the particle can
be charged from a completely depleted state. Note that
a reverse reaction of order greater than one can be considered but
this is not usually used.

Such a local behaviour of the Butler-Volmer condition is included in very few formulations, see \cite{doyle93} and \cite{west}, and, without explanation of why this choice is taken. A much greater number of articles in the literature do not include this local behaviour and, also, use a formulation of $U_{\rm eq}$ independent of $c$. These are not well-suited for extremely low electrolyte concentrations. Nevertheless, it is useful to model these cases since, for example, during high discharge rates, the electrolyte can be almost depleted of lithium in some parts of the cell (see \cite{bizeray2015lithium}).

One reasonable way to rewrite the Butler-Volmer relationship that emphasises the forward and reverse parts of the reactions is to introduce two strictly positive bounded functions {$i_a(c_s)$}, and $i_c(c_s,c)$ where
\be \label{fdi0}
i_0 (c_{\rm s},c)&=&
i_{\rm a}(c_{\rm s})^{\frac{\alpha_{\rm c}}{\alpha_{\rm a}+\alpha_{\rm c}}}
i_{\rm c}(c,c_{\rm s})^{\frac{\alpha_{\rm a}}{\alpha_{\rm a}+\alpha_{\rm c}}}\\
\label{fdU}
\mbox{and} \qquad U_{\rm eq} (c_{\rm s},c)&=&
 \frac{RT}{F(\alpha_{\rm a}+\alpha_{\rm c})}\log \left( \frac{i_{\rm c}{(c_{\rm s},c)}}{ i_{\rm a}(c_{\rm s})} \right) {+ u_{\rm eq} (c_{\rm s},c)},
\ee
{where $u_{\rm eq}$ is a bounded function, } so that the Butler-Volmer equation (\ref{DBV}) for the transfer current can be written in the form
\be \begin{split}
j_{\rm tr} ={i_{\rm a}}(c_s)\exp\left[\frac{\alpha_{\rm a} F}{RT}(\phi_s - \vph  {- u_{\rm eq} (c_{\rm s},c)})\right] \\ -{i_{\rm c}}(c_s,c)\exp\left[\frac{-\alpha_{\rm c} F}{RT}(\phi_s - \vph  {- u_{\rm eq} (c_{\rm s},c)})\right] .
\label{starr}
\end{split} \ee
The functions $i_{\rm a}$ and $i_{\rm c}$ must be strictly positive except at the extremes of the concentrations and locally these functions must have the following behaviours:
\be
{i_{\rm a}(c_{\rm s})} \sim  k_{\rm a} \;c_{\rm s} \quad \mbox{as} \quad c_{\rm s} \to 0 ,\\
i_{\rm c} (c_{\rm s},c) \sim k_{\rm c,s}(c)\; (c_{\rm s,max}-c_{\rm s}) \quad \mbox{as} \quad c_{\rm s} \to c_{\rm s,max},\\
i_{\rm c} (c_{\rm s},c) \sim  k_{\rm c,e}(c_{\rm s})\; c \quad \mbox{as} \quad c \to 0,
\ee
with $k_{c,s}(c)$ and $k_{c,e}(c_s)$ strictly positive functions and $k_{\rm a}$ a positive constant, all of which may differ in different electrodes. Note there may also be a need to consider a maximum electrolyte concentration to avoid precipitation and this might be accommodated in a similar manner.

Notice four important features of the approach proposed here, in
contrast with what is commonly found in the literature:
\begin{enumerate}
\item The function $U_{\rm eq}$, the OCP,
 also depends on the electrolyte concentration $c$ (instead
  of being independent of it).
 \item If $c_{\rm s}\rightarrow 0$, with $c\neq 0$, then $i_0\rightarrow 0$ and $U_{\rm eq}\rightarrow \infty$ (instead of tending to a finite value) so that
 \begin{equation}\begin{split} \label{ssssdf1}
j_{\rm tr} \sim
k_{\rm a}  c_{\rm s} \exp \left( \frac{\alpha_{\rm a}F}{R \ T}(\phi_{\rm s}-\vph {- u_{\rm eq} (0,c)})) \right)\\ -i_{\rm c}(0,c)\exp \left( \frac{-\alpha_{\rm c}F}{R \ T}(\phi_{\rm s}-\vph {- u_{\rm eq} (c_{\rm s},c)}))
\right).
\end{split}  \end{equation}
  \item If $c\rightarrow 0$, with $c_{\rm s}\neq 0$, then $i_0\rightarrow 0$ and $U_{\rm eq}\rightarrow -\infty$ so that
 \begin{equation} \begin{split} \label{ssssdf2}
j_{\rm tr} \sim
i_{\rm a} (c_{\rm s}) \exp \left( \frac{\alpha_{\rm a}F}{R \ T}(\phi_{\rm s}-\vph {- u_{\rm eq} (c_{\rm s},0)})) \right)
\\ -k_{\rm c,e}(c_{\rm s})\; c \exp \left( \frac{-\alpha_{\rm c}F}{R \ T}(\phi_{\rm s}-\vph {- u_{\rm eq} (c_{\rm s},0)}))
\right) ,
\end{split}  \end{equation}
\item
If $c_{\rm s}\rightarrow c_{\rm s,max}$, with $c\neq 0$, then $i_0\rightarrow 0$ and $U_{\rm eq}\rightarrow -\infty$
and
\begin{equation}\begin{split} \label{ssssdf3}
j_{\rm tr} \sim i_{\rm a} (c_{\rm s,max}) \exp \left( \frac{\alpha_{\rm a}F}{R \ T}(\phi_{\rm s}-\vph {- u_{\rm eq} (c_{\rm s,max},c)})) \right) \\ - k_{\rm c,s}(c) (c_{\rm s}-c_{\rm s,max})\exp \left( \frac{-\alpha_{\rm c}F}{R \ T}(\phi_{\rm s}-\vph {- u_{\rm eq} (c_{\rm s,max},c)}))
\right).
\end{split} \end{equation}
\end{enumerate}
These conditions ensure that the lithium flux is constrained to
prevent concentrations in the solid being taken into nonphysical
regimes and also ensure that, if the solid is nearly fully depleted (or filled) with lithium, that the transfer current is not artificially  forced to zero. Thus a Butler-Volmer relation of this form allows the flux to move the system away from these depleted (and filled) states. This gives a  model which is capable of  describing battery charge from a completely depleted state or discharge
from a fully charged state.

The simplest functions $i_0$ and $U_{\rm eq}$ {that we can use in the Butler-Volmer Equation (\ref{DBV}) that satisfy} the above conditions, have the form
\begin{equation}
i_{\rm a}{(c_{\rm s})} = k_{\rm a} \; c_{\rm s}
\end{equation}
\begin{equation}
i_{\rm c}(c_{\rm s},c) =
  k_{\rm c} \; c\;(c_{\rm s,max}-c_{\rm s}),
\end{equation}
and in this instance, according to (\ref{fdi0}) and (\ref{fdU}),
$$
 U_{\rm eq}(c_{\rm s},c)
=
 \frac{RT}{F(\alpha_{\rm a}+\alpha_{\rm c})}\log \left( \frac{k_{\rm c}\; c\;(c_{\rm s,max}-c_{\rm s})}{ k_{\rm a}\; c_{\rm s}} \right) + u_{\rm eq}(c_s,c),
$$
and
\be
i_0 (c_{\rm s},c) &=&
  k \; c^{\frac{\alpha_{\rm a}}{\alpha_{\rm a}+\alpha_{\rm c}}} \; (c_{\rm s,max}-c_{\rm s})^{\frac{\alpha_{\rm a}}{\alpha_{\rm a}+\alpha_{\rm c}}} \; c_{\rm s}^{\frac{\alpha_{\rm c}}{\alpha_{\rm a}+\alpha_{\rm c}}} , \label{necpi0}
\ee  
{where 
\begin{equation} \label{3ks}
k=k_{\rm a}^{\frac{\alpha_{\rm c}}{\alpha_{\rm a}+\alpha_{\rm c}}} k_{\rm c}^{\frac{\alpha_{\rm a}}{\alpha_{\rm a}+\alpha_{\rm c}}},
\end{equation}}
which coincides with the commonly used relation (\ref{ddio}), for $i_0 (c_{\rm s},c)$, where $\alpha_{\rm a}+\alpha_{\rm c} =1$ (note in almost all the literature $\alpha_a=\alpha_c=0.5$).

Some authors
(cf. \cite{2006SmWa,2006-b-SmWa,2007SmRaWa,2012KiSmIrPe})) take a
constant value for $i_0$ and some of them (cf. \cite{2006-b-SmWa})
claim that it exhibits modest dependency on electrolyte and solid
surface concentration. Although this can be valid for appropriate
particular constrained cases, in a general situation this is
inappropriate since $i_0$ should vary near the extreme cases when
the battery is either close to being fully charged or fully discharged.

\subsubsection{Practical methods of fitting data}

When fitting the various functions characterising the Butler-Volmer equation (\ref{DBV}) to the data {it is necessary to include the singular behaviours as $c\rightarrow 0$,  $c_{\rm s}\rightarrow 0$ and $c_{\rm s}\rightarrow c_{\rm s,max}$ as given previously. A} sensible approach {is to make non-singular modifications to the simple model (\ref{necpi0}) as follows}. 
There is usually very little accurate data on the exchange current $i_0$ and so (\ref{necpi0})--(\ref{3ks}) is typically the form that is taken. For the OCP one approach is to take
\begin{equation}
\begin{array}{l}
{\displaystyle U_{\rm eq}(c_{\rm s},c) = \frac{RT}{F(\alpha_{\rm a}+\alpha_{\rm c})}\log \left( \frac{{k_{\rm c}}\;c\;(c_{\rm s,max}-c_{\rm s})}{ {k_{\rm a}}\;c_{\rm s}} \right) + f(c_{\rm s}; a_1, a_2,\cdots , a_N) }
 \\[.3cm] \hspace{3cm} + g(c; b_1,\cdots , b_M),
\end{array}
\end{equation}
where $f(\cdot ; a_1, a_2,\cdots , a_N)$ is some family of suitably {bounded} functions, and  $g$ is a similar function allowing for changes in $c$, while ${k_{\rm c}}, {k_{\rm a}}, a_1,\cdots a_N${, }$b_1,\cdots b_M$ {is a set} of parameters to be chosen by fitting experimental data{, with $k, k_{\rm c}, k_{\rm a}$ satisfying (\ref{3ks})}. Note this could also be expressed in terms of the non-dimensional lithium stoichiometry $y= {c_{\rm s}}/{c_{\rm s,max}}$

{For} example, using the function $f$ proposed in \cite{2014WeSuPe} and a polynomial function for changes in $c$, the OCP can be fitted to 
\begin{equation}\label{dcd9}
\begin{array}{lll}
U(c_{\rm s},c) & = & {\displaystyle \frac{RT}{F(\alpha_{\rm a}+\alpha_{\rm c})}\log \left( \frac{{k_{\rm c}}\;c(c_{\rm s,max}-c_{\rm s})}{ {k_{\rm a}}\; c_{\rm s}} \right)
+ a_0} \\
& &  {\displaystyle + a_1\frac{1}{1+{\rm e}^{\alpha_1(\frac{c_{\rm s}}{c_{\rm s,max}}-\beta_1)}} + a_2\frac{1}{1+{\rm e}^{\alpha_2(\frac{c_{\rm s}}{c_{\rm s,max}}-\beta_2)}}
+ a_3\frac{1}{1+{\rm e}^{\alpha_3(\frac{c_{\rm s}}{c_{\rm s,max}}-1)}} } \\
& &  {\displaystyle  + a_4\frac{1}{1+{\rm e}^{\alpha_4\frac{c_{\rm s}}{c_{\rm s,max}}}} + a_5 \frac{c_{\rm s}}{c_{\rm s,max}}  + \sum_{i=1}^Mb_i c^i}.
\end{array}
\end{equation}
For practical {particular cases,} because there is usually very little data related to variations with $c$, it may be appropriate to take $b_i \equiv 0$, so
that the only $c$ dependency is in the logarithm and {this} will still ensure the
extreme behaviour near $c=0$ is physical.


\subsection{Properties of common electrode materials} \label{common}
The depictions of lithium transport in the active material of the electrode (electrode particles) in \S \ref{litrans}, and of the charge transfer reaction between the active material and the electrolyte in \S \ref{bv}, although commonly employed, are an oversimplification of the true behaviour of these materials. In order to highlight some of the nuances of modelling these materials, we briefly review a small part of the copious literature, focusing on some commonly used materials.

The standard negative electrode (anode) material used in commercial lithium-ion batteries is graphitic carbon which alloys with lithium to form Li{$_y$}C$_6$, see,   \cite{persson10}. Other materials (such as silicon) are currently being developed with the aim of supercedeing graphite in this role in the future but none has reached the stage of commercialisation. In contrast to the situation for negative electrodes, there are a wide range of positive electrode materials currently used in commercial batteries; these are reviewed in \cite{julien14}.

\cite{julien14} notes that most electrode materials fall into three categories, based on the lithium ion diffusion pathways in the material. In spinel materials lithium transport is three-dimensional, in layered materials it is predominantly two-dimensional and in olivines it is predominantly one-dimensional. Commonly used commercial positive electrode materials include LiMn$_2$O$_4$ (often referred to as LMO) which is a spinel, {NMC (mentioned above)} which is a layered material and LiFePO$_4$ (often referred to as LFP) which is an olivine. Li{$_y$}C$_6$ (the standard negative electrode material) has a layered structure. The dimensionality of lithium transport in layered (2-d) and olivine (1-d) materials suggests that use of an isotropic model of lithium transport, such as (\ref{lindiff}), is inadequate and should, at the very least, be generalised to an anisotropic diffusion model capable of capturing, for example, the dependence of transport properties on alignment with the crystal axes. Nevertheless an isotropic transport model is frequently used even when it might seem inappropriate. For example, \cite{srinivasan04} and \cite{arora00,doyle96} use isotropic diffusion models for lithium transport in LFP (an olivine) electrode particles and graphitic carbon (a layered material) electrode particles, respectively. There may, however, be good reasons for doing this; for example electrode particles are rarely formed from single crystals and lithium transport in conglomerate particles, formed from randomly oriented crystals, might reasonably be expected to appear isotropic on the lengthscales of interest.

\subsubsection{Lithium transport in Li{$_y$}C$_6$ (negative electrode material).}
Much of the early modelling work on lithium-ion batteries, such as \cite{arora00,doyle96,fuller94,fuller94b}, modelled lithium transport within Li{$_y$}C$_6$ electrode particles by a linear diffusion equation (\ref{lindiff}). However both \cite{takami95,verbrugge03} and \cite{krach18} suggest a very strong dependence of solid-state lithium diffusion coefficient $D_s(c_s)$ with lithium concentration $c_s$ in Li{$_y$}C$_6$ particles, with a range of variation of up to about  two orders of magnitude, depending upon the exact form of carbon used. In both sets of experiments the size of the carbon particles used was around 10$\mu$m and diffusion decreased markedly as lithium concentration $c_s$ was increased. To complicate matters further Li{$_y$}C$_6$ is known to exhibit (at least) three phases as lithium {stoichiometry $y$} is increased. The presence of these phases can be seen inferred from colour changes to the electrode particles (dark blue--low lithium, red--intermediate lithium and gold--high lithium). In \cite{harris10} optical microscopy measurements are used to characterise the phase transitions occurring (with increasing lithiation) within a Li{$_y$}C$_6$ half-cell anode subject to uniform charging. This shows different phases co-existing (in distinct graphite electrode particles) at different positions in the anode. However, the relevance of these results to commercial devices should perhaps not be overstated, because the width of the negative electrode used in \cite{harris10} is particularly large (around 800$\mu$m) compare to the standard electrode size in commercial devices (around 100$\mu$m). For this reason, the charging process observed in \cite{harris10} is likely to be limited by lithium diffusion within the electrolyte, as the electrode particles deplete the surrounding electrolyte of lithium ions. \cite{Tho17} have also applied optical microscopy to graphite electrodes and observed considerable spatial nonuniformity even after the electrode was left quiescent for an extended period. They were able to predict such states by employing a Cahn-Hilliard phase field model which will be discussed further in \S\ref{cahn}.

Graphite reacts with the electrolyte, consuming lithium ions, to form a thin layer of solid material on the graphite which is referred to as the  solid-electrolyte interphase (SEI) layer and, as noted  in \cite{bruce08}, is essential for maintaining the structural integrity of the electrode particles. However, the consumption of electrolyte by this reaction means that graphite electrode particle size cannot be reduced to the nanoscale (in an attempt to improve the charge/discharge rate of the battery) without severely compromising battery capacity \cite{bruce08}; some lithium makes up the SEI layer and can no longer participate in the useful reactions that store charge. This SEI layer also forms a barrier to lithium ion (and current) transfer between the electrolyte and electrode which may have a significant bearing on electrode performance and has thus been incorporated into some models, for example \cite{srinivasan04b}. Diffusion of lithium along the graphene sheets of pure single crystal graphite is extremely fast \cite{persson10} (diffusion coefficient of the order of $10^{-7}-10^{-6}$cm$^2$s$^{-1}$), so that, even in electrodes comprised of quite large electrode particles ($\sim 100 \mu$m) discharged (or charged) at very high rates, it should not significantly affect cell performance. However, \cite{persson10} demonstrates that diffusion perpendicular to the graphene sheets and along grain boundaries is many orders of magnitude slower (diffusion coefficient of the order of $10^{-11}$cm$^2$s$^{-1}$) and uses this to infer that this high degree of anisotropy can be used to explain the widely disparate measurements of diffusivity reported in polycrystalline graphite.

\subsubsection{Lithium transport in NMC and LMO (positive electrode materials).}
Lithium diffusion in NMC \cite{wu12} is highly nonlinear so that  $D_s(c_s)$ decreasing by about two orders of magnitude as lithium concentration within the material increases. Furthermore NMC can be charged and discharged at high rates and the OCP is smooth without the stepped plateau features that usually characterise phase transitions. In contrast, \cite{julien14} notes that LMO undergoes a number of phase transitions as it charges and discharges, which are associated with plateaus in its OCP curve. Diffusivity of lithium in single crystals of LMO is about an order of magnitude lower than in multicrystalline particles \cite{das05} suggesting that grain boundaries form an easy pathway for lithium diffusion. Furthermore LMO has the disadvantage of capacity loss and fade after repeated cell cycling. This capacity fade has been ascribed, by \cite{das05}, to the formation of a SEI layer, and consequent loss of lithium mobility.

\subsubsection{Lithium transport in LFP (positive electrode
  material).} 
\cite{bruce08} point out that intercalation in LFP involves a phase transition between FePO$_4$ and
LiFePO$_4$, which is reflected in its flat 
OCP curve (see Figure \ref{equilibrium_U}(b)). \cite{Kang09} note that the transport of
lithium is dominated by transport along channels in particular
crystalline directions, the $b$-direction, and in single
crystal nanoparticles is extremely rapid, so fast indeed
that it is doubtful that lithium intercalation in LFP single crystal
nanoparticles will ever limit battery performance. This point is
clearly made by \cite{johns09}, who demonstrate that
discharge in a half cell nanoparticulate LFP cathode is limited by
conduction and transport in the electrolyte.  In larger LFP electrode
particles, \cite{jugovic09} point out that
performance is significantly impaired because lithium ion transport
along the b-direction channels is easily obstructed by grain
boundaries and crystal defects. This gives rise to an apparent lithium
diffusivity in LFP that decreases sharply as the size of the particle
increases, see \cite{malik10}. Standard models of this material include the
so-called shrinking core model, presented in \cite{srinivasan04}, which attempts to
capture the phase transition by using a one-dimensional free-boundary
model of the phase transition (akin to a Stefan model, see \eg \cite{Ruben}) in a
spherically symmetric electrode particle. This approach is used in a
3D-scale model, that captures agglomeration of LFP nanocrystals into
agglomerate particles, by \cite{dargaville10} who
show good agreement with experimental discharge curves for a wide
range of discharge rates. However the shrinking core model is known to
predict distributions of the two-phases that are not observed in
practice and it is also not easy to implement in a form that allows
numerous charge-discharge cycle because of the appearance of multiple
free-boundaries, as noted by \cite{farkhondeh11}. A simpler alternative, suggested
in \cite{farkhondeh11}, is to model lithium transport within LFP
particles by a phenomenological nonlinear diffusivity $D_s(c_s)$ and
this appear to fit discharge data well.


\subsection{An approach based on Cahn-Hilliard equations of phase separation \label{cahn}}
The lack of an entirely satisfactory theory of lithium transport in
electrode materials that exhibit phase transitions (such as graphite
and LFP) has recently led to an alternative, and more fundamental approach in which phase separation with the electrode material is
modelled using a Cahn-Hilliard equation. The first use of this
approach in this context was by \cite{han04}, who used it to
simulate a generic two-phase material. Subsequently
\cite{bai11,cogswell13,zeng14} and \cite{singh08} applied this method to LFP using it to
study phase separation (and its suppression) in LFP nanoparticles.
Both \cite{ferguson12} and \cite{dargaville13} have incorporated a
Cahn-Hilliard based phase-field description of lithium transport
within LFP electrode particles into a porous electrode model. Notably
\cite{dargaville13} compare their results to
experimental discharge curves over a wide range of discharge rates,
but are unable to obtain a particularly good match to data. In \cite{zeng14} it is observed that the model is sufficient to capture transitions from solid-solution radial diffusion
to two-phase shrinking-core dynamics. In
\cite{ferguson14} fit to data from both LFP and
graphite half cells, at very slow discharge rates, with some degree of
success (particularly in predicting the positions of the phase
transitions across the graphite electrode). One remarkable feature of
this work is that it predicts the observed steps in the 
OCP curves, as a consequence of the phase transitions rather than having to fit a
stepped OCP to a potential function as in the standard Newman type
model.  However these Cahn-Hilliard type models are probably only
directly applicable to small single crystal electrode particles,
because of the extra physics required to model, for example,
obstruction of lithium transport by grain boundaries and defects in
larger particles. As mentioned previously in practical applications where the crystals are very small they usually do not limit battery discharge and so accurately capturing their internal transport may be of secondar importance in predicting cell-level features.


 \section{A coupled device scale model \label{device}}

 The aim of this section is to discuss how macroscopic device-scale
 equations can be systematically derived from a model of the
 electrolyte surrounding the electrode particles, the geometry of
 the electrode particles and a description of the electrolyte
 reactions taking place on the surface of the electrode particles. In
 this we follow the work of \cite{richardson12} who
 derived the device scale equations for an ideal (dilute) electrolyte
 and \cite{ciucci11} who derived the device scale
 equations from a model of a moderately-concentrated electrolyte. We
 remark that the moderately-concentrated electrolyte model used by
 \cite{ciucci11} predicts that electrolyte conductivity $\kappa(c)$ is
 proportional to the ionic concentration $c$ (as it would for an ideal
 solution) and so is incapable of adequately describing electrolytes
 at the concentrations typically occurring in a commercial lithium ion
 cell. Here we shall extend the work in \cite{richardson12} to the
 moderately-concentrated solution model described in \S \ref{conc} and
 which is applicable to most battery electrolytes.


\subsection{The microscopic model \label{micro}}
The purpose of this section is to set out the equations and boundary
conditions of a detailed microscopic model of the battery electrode,
including both lithium transport and current flow through the
electrode particles and the electrolyte. A portion of a typical
electrode geometry is illustrated in Figure \ref{cell_struct}(a), in
which a periodic array of electrode particles occupying region $\OP$
(here they have ellipsoidal shape as a
possible example) is surrounded by the electrolyte, which occupies the
region $\VP$, and the interface is $\dd \OP$. As will be discussed further in \S\ref{binsec} we will assume that the volume occupied by the binder and conductive filler is negligibly small so that the electrode particles and electrolyte completely fill the electrode.

Charge transport in the electrolyte is described by the equations (\ref{mcon4})-(\ref{mcon6}). At the interface with an electrode particle the transfer current density is given by the Butler-Volmer relation (\ref{DBV}) and this can be equated to the current flowing into the electrolyte via the boundary condition
\be
\jj \cdot \vect{N}|_{\dd \OP} = j_{\rm tr} (c,\vph,c_s,\phi_s)|_{\dd \OP}, \label{mcon7}
\ee
where $\vect{N}$ is the unit outward normal to the interface (it points into the electrolyte region).
{A boundary condition for (\ref{mcon4}), the equation for conservation of lithium ions, is provided by noting that
all charge transfer across this surface takes place via the motion of lithium ions. Hence, we have}
the following condition on $\vect{q}_p$, the flux of positively charged lithium ions,
\be
\q_p\cdot \vect{N}|_{\dd \OP} = \frac{1}{F} j_{\rm tr}(c,\vph,c_s,\phi_s)|_{\dd \OP}, \quad \mbox{where} \quad \q_p=-\de(c) \nabla c+\frac{\tn}{F} \jj.  \label{mcon8}
\ee
Note that (\ref{mcon7}) and  (\ref{mcon8}) imply that the flux of negative ions is zero with $\q_n\cdot \vect{N}|_{\dd \OP}=0$, as required physically.

In each individual electrode particle $\OP$, a diffusion equation is solved for lithium concentration in the active material. Generalising (\ref{lindiff}) to an arbitrary shaped particle and allowing for nonlinear diffusion gives
\be
\frac{\dd c_s}{\dd t}= \nabla \cdot ( D_s(c_s) \nabla c_s) \qquad \mbox{in} \quad \OP, \label{nixon}
\ee
with boundary condition
\be
-D_s(c_s) \nabla c_s \cdot \vect{N}|_{\dd \OP}= \frac{1}{F}  j_{\rm tr}(c,\vph,c_s,\phi_s)|_{\dd \OP}. \label{spiro}
\ee


\subsection{Homogenising the equations in a porous electrode\label{homog}}
Here we homogenise the moderately-concentrated electrolyte equations (\ref{mcon4})-(\ref{mcon6}), with boundary conditions (\ref{mcon7})-(\ref{mcon8}), over a porous electrode formed by an array of electrode particles permeated by the electrolyte. In order to do this we assume that the electrode can be subdivided into an array of cells over which the electrode structure is {\it locally} periodic; that is, the structure inside neighbouring cells is virtually identical but may differ significantly between cells separated on the macroscopic lengthscale.\footnote{In order to use the method of homogenisation it has been shown that cell structure need not be entirely periodic, only almost so on the microscopic lengthscale, \cite{richardson11}.} An example of cell microstructure, around an array of ellipsoidal electrode particles, is illustrated in Figure \ref{cell_struct}. 
\begin{figure}
\begin{center}
\includegraphics[width=0.6\columnwidth, height=0.3\columnwidth]{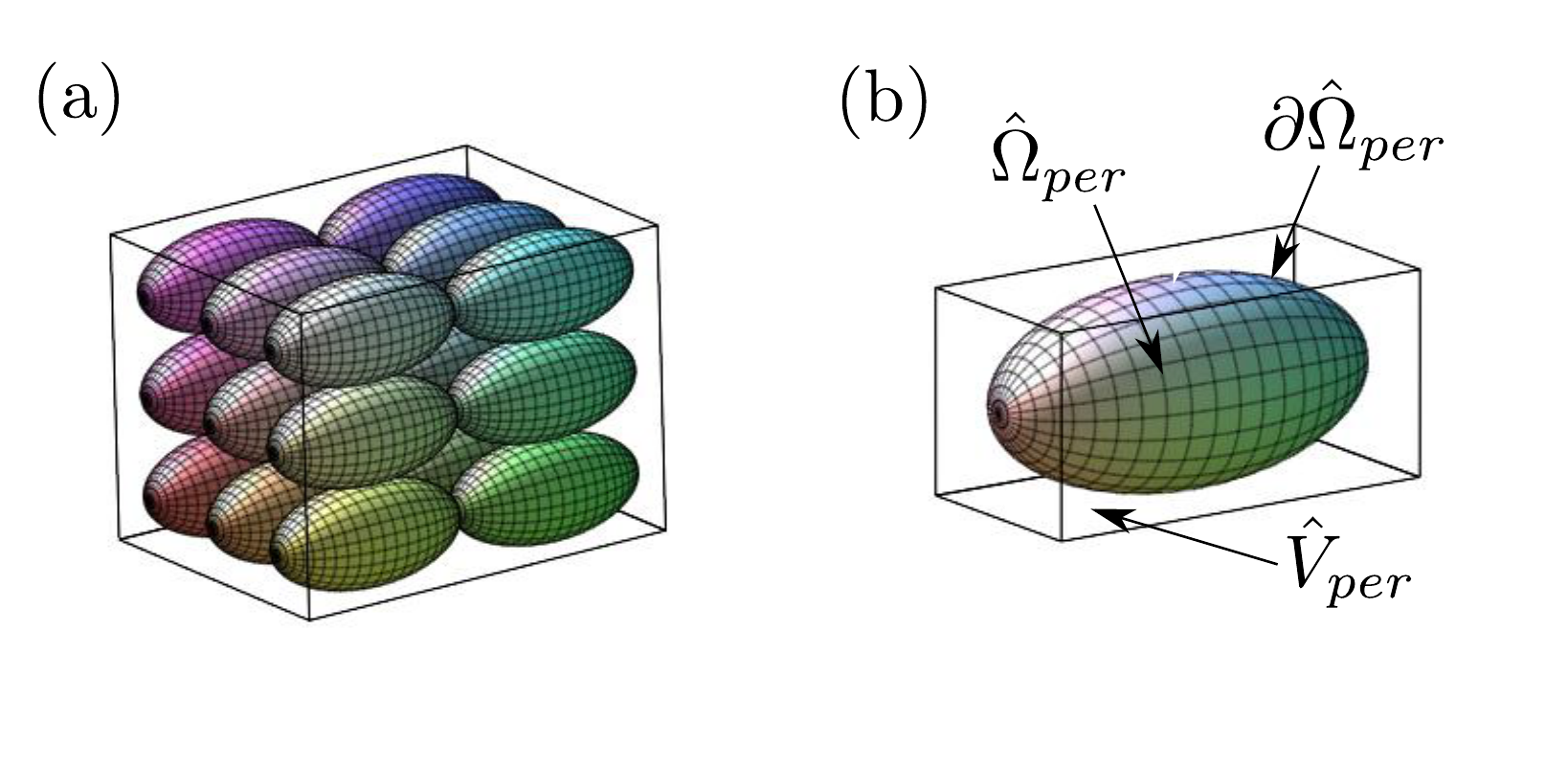}
\caption{(a) An example of a periodic microstructure with ellipsoidal electrode particles. (b) An illustration of the microstructure geometry within a periodic cell $\VP \cup \OP$, about an individual electrode particle.}
\label{cell_struct}
\end{center}
\end{figure}
To average the problem using homogenisation we introduce a variable $\hat{\bf{x}}$ to indicate position in the microscopic cell and another variable $\xx$ to indicate macroscopic position in the entire electrode. 
Since a very similar analysis has been conducted in \cite{richardson12} for a dilute electrolyte we omit the details of the analysis here and merely write down the results (in dimensional form). These consist of macroscopic equations for the lithium ion concentration $c$ and the electrolyte potential (measured with respect to a lithium electrode) $\vph$ that are formulated in terms of the microscopically volume averaged lithium ion flux $\avqp$, the microscopically volume averaged current density $\avj$ and the microscopically surface averaged transfer current density $\avjn$. These averaged quantities are formally defined in terms of integrals over the microscopic cells as follows:
\be
\avj=\frac{1}{\mv+\mo} \int_{\VP} \jj\; \dV, \qquad \avqp=\frac{1}{\mv+\mo} \int_{\VP} \vect{q}_p\; \dV, \\
 \avjn=\frac{1}{\so} \int_{{\dd} \OP} j_{{\rm tr}}\; \dS,
~~~~~~~~~~~~~~~~~~~~~~~~~~~~~
\label{avjn1}
\ee
where the integrals are over the microscopic variable and the averaged functions depend only on the macroscopic space variable $\xx$ 
and time $t$.  The quantities $\mv(\x)$, $\mo(\x)$ and $\so(\x)$ are defined by
\be
\mv= \int_{\VP} \dV, \qquad \mo=\int_{\OP} \dV, \qquad  \so=\int_{\dd \OP} \dS
\ee
and respectively give the volume of electrolyte, the volume of electrode particle and the surface area of electrode particle in the microscopic cell. The macroscopic homogenised electrolyte equations then take the form
\be
\ev \frac{\dd c}{\dd t} + \nabla \cdot  \avqp= \frac{\bet}{F} \avjn,  \qquad \avqp= -\de(c) \B \nabla c + \frac{\tn}{F} \avj, 
\label{presi}\\
\nabla \cdot  \avj= \bet \avjn,  \qquad \avj=-\kappa(c) \B \left( \nabla \vph - 2 (1-\tn) \frac{RT}{F} { \frac{a_e'(c)}{a_e(c)}\nabla c }\right),
\label{dent}
\ee
{where the final term in the current equation \eqref{dent} takes the standard form $2 (1-\tn) ({RT}/{F c}) \nabla c$ when the activity is that for an ideal solution (\ie $a_e(c)=c/c_T$).} It is interesting to note that the equations for the electrolyte after homogenisation, (\ref{presi})--(\ref{dent}), are very similar in nature to the original equations for the pure electrolyte, (\ref{mcon4})--(\ref{mcon6}), except that there are now source terms in the conservation equations corresponding to the transfer current from the electrodes. Here $\epsilon_v$ is the volume fraction of the electrolyte defined  by
\bes
\epsilon_v= \frac{|\Vnh|}{|\Vnh|+|\OPnh|},
\ees
the Brunauer-Emmett-Teller surface area (BET surface area), $b_{et}$, \ie the surface area of particles per unit volume of electrode, is defined by
\bes
b_{et}= \frac{\int_{\OPnh} {\rm d}S}{|\Vnh|+|\OPnh|},
\ees
and $\B$ is the dimensionless permeability tensor whose nine components are defined by the relations
\be \begin{split}
B_{ij} = \frac{1}{\mv+\mo} \int_{\VP}  \left( \delta_{ij} - \frac{\dd \chi^{(j)}}{\dd x_i} \right) \dV \\ \mbox{for} \quad i=1,2,3 \quad  \mbox{and} \quad j=1,2,3,
\end{split} \ee
in which the three characteristic functions $\chi^{(j)}$ ($j=1,2,3$) are solutions to the local cell problems
\be
  \left.
    \begin{array}{c}
      \hat{\nabla}^2 \chi^{(j)}=0   \quad \mbox{in} \quad \VP, \\
      \hat{\nabla} \chi^{(j)} \cdot \nn |_{\dd \OP }= \vect{e}_j \cdot \nn |_{\dd \OP }, \\
      \chi^{(j)} \quad \mbox{periodic in $\hat{x}$ on} \quad \VP, \\
      \int_{\VP} \chi^{(j)}\; \dV=0
    \end{array} \right\} \quad \mbox{for}  \quad i=1,2,3,
\label{chiprob}
\ee
where $\vect{e}_j$ is a basis vector in the $\hat{x}_j$-direction and $\nn$ is the unit outward normal (pointing from $\OP$ into $\VP$) to the surface $\dd \OP$.

We note also here the possibility of using this type of homogenisation technique in conjunction with microscale three-dimensional image data obtained from real battery electrodes in order to obtain more realistic representations of the geometric parameters $\epsilon_v$, $B_{ij}$ and $b_{et}$, as discussed for example in \cite{gully14} and \cite{foster15}. In many papers the tensor $\B$ is taken to be a constant times the unit tensor, which correspond to a highly symmetric set of particles, such as spherical particles on a regular lattice, and proves to be quite a reasonable model.

The homogenised electrolyte equations (\ref{presi})-(\ref{dent}) must be solved in conjunction with the macroscopic equations for the current flow through the solid part of the electrode. These can be obtained by using a constitutive law for current flow in the electrode matrix (\ref{ohm-solid}) with a current conservation equation that accounts for transfer of charge from the electrode matrix into the electrolyte
\be
\nabla \cdot \jj_s= -\bet \avjn, \quad \mbox{where} \quad \jj_s= -\kappa_s \nabla \phi_s. \label{ford}
\ee
The system of macroscopic equations (\ref{presi})-(\ref{dent}) and
(\ref{ford}) require to be solved along with the microscopic
lithium transport equations (\ref{nixon}) and (\ref{spiro}), at each
point in macroscopic space, in order to determine
$c_s|_{\dd \OP}$, which is required to obtain the transfer
current $j_{\rm tr}(\phi_s-\vph,c_s,c)$ (and hence $\avjn$ 
as given in (\ref{avjn1})).  

{Note that where the electrode particles are spherical (and isotropic) $c_s|_{\dd \OP}$ is uniform over the particle surface and is thus just a function of the macroscopic variables. It follows therefore that $j_{\rm tr}$ also just a function of the macroscopic variables and, as a consequence of the averaging equation \eqref{avjn1} it follows that for 
\be 
\avjn=j_{\rm tr} \quad \mbox{for spherically symmetric electrode particles.} \label{sphsimp}
\ee}

\subsubsection{\label{binsec}Remarks on the role of binder.}
{In the discussion above we have assumed that the electrode was formed solely from electrode particles bathed in electrolyte. While this is often a reasonable description of research cells, many commercial devices also incorporate a significant volume fraction of polymer binder material that acts both to enhance the structural integrity of the device and, in combination with a conductivity enhancer (such as carbon black), maintain good electrical contact between electrode particles (these are often poor conductors). Three dimensional images of typical commercial electrodes using focused ion beam in combination with scanning electron microscopy (FIB-SEM) can be found in \cite{gully14,foster15} and \cite{liu16}. These show a porous binder material filling almost all the space between electrode particles with the exception of some linear features around the electrode particles where it appears that the binder has become delaminated from the electrode particles.\footnote{A physical explanation for this binder delamination is provided in {\cite{foster16,foster17}}.} The porosity and pore size of the binder materials varies significantly between different electrode types. Typical pore sizes are usually in the range 10-500nm, much smaller than typical electrode particle sizes which are usually at least micron sized. 
To account for these effects within an homogenisation approach the analysis could be modified in two possible ways. 
Firstly it could be performed directly on a microstructure in which all three constituents (electrode particle, electrolyte and binder) are resolved by the cell problem and would follow a very similar pattern to that described above in which the binder were treated as part of $\OP$ with an interface with the electrolyte on which the transfer current density $j_{\rm tr} \equiv 0$. 
Secondly, because obtaining a good representation of the pore geometry in the binder is challenging, even using modern high-performance microscopy, see \cite{liu16}, it is probably better to treat
 the electrolyte permeated nanoporous binder as a single electrolyte material (albeit it one with reduced electrolyte volume fraction, electrolyte diffusivity and conductivity) and homogenise over the electrode particles and this composite material. Indeed this second approach
 is a standard way of treating such two phase (electrolyte/binder) materials in the literature which are often termed porous solid polymer electrolytes (see, \eg \cite{miao08}).}

\subsubsection{The pseudo 2d-model.\label{p2dsec}} 
As mentioned previously a common approach in the literature to modelling practical batteries is to use the so-called {\em pseudo-2d model} where the
behaviour on the macroscale is one-dimensional {(with spatial position denoted by $x$) and transport of lithium on the microscale takes place within spherical particles, and is thus also one-dimensional taking place in the particles' radial direction (with radial position denoted by $r$).} These assumptions are tantamount to assuming that the averaged macroscopic vector-valued currents and fluxes from \S\ref{micro} and \ref{homog} are only non-zero in the $x$-direction. Henceforth we will replace these vector quantities with scalar counterparts. Here we describe the system of equations that arise for such a situation exploiting the homogenisation results described previously {in \eqref{presi}-\eqref{ford}} and including the typical units.

As alluded to above, $x$ denotes position across the cell in the direction perpendicular to the current collectors which are positioned at $x=L_1$ and $x=L_4$ respectively, so that $x \in (L_1,L_4)$. The cell is subdivided into three regions (as illustrated in figure \ref{geom_sketch}) with the negative electrode occupying the region $x \in (L_1,L_{2})$, the separator occupying the region $x \in (L_2,L_3)$ and the positive electrode occupying the region $x\in (L_3,L_4)$. In the electrolyte, which permeates the whole cell (\ie $x\in (L_1,L_4)$), we seek to determine the Li concentration $c(x,t)$ (mol m$^{-3}$), the electric potential (measured with respect to a reference lithium electrode) $\vph (x,t)$ (V) and the ionic current density $\langle j \rangle(x,t)$ (A\,m$^{-2}$). In the negative electrode ($x \in (L_1,L_2)$) we seek solutions for the solid phase potential $\phisa(x,t)$ and the solid phase current $\jsa(x,t)$. In addition we seek to determine the lithium distribution $\csa(r,x,t)$ within the (negative) electrode particles (of radius $\Ra(x)$) as a function of position $r \in [0,\Ra(x)]$ within the particle and the position $x$ of the particle within the electrode. Similarly in the positive electrode ($x \in (L_3,L_4)$) we seek solutions for the solid phase potential $\phisc(x,t)$, the solid phase current $\jsc(x,t)$ and also  the lithium distribution $\cscc(r,x,t)$ within the (positive) electrode particles (of radius $\Rc(x)$) as a function both of position $r\in [0,\Rc(x)]$ within the particle and the position $x$ of the particle within the electrode. A sketch of the device geometry as well as an illustration of the domains of definition of the dependent variables is shown in figure \ref{geom_sketch}. Throughout what follows we assume that the transport of lithium within both negative and positive electrode particles occurs through nonlinear isotropic diffusion, though as discussed in \S \ref{common} this is not the only possibility. Furthermore since the electrode particles are spherical we can make use of the simplification \eqref{sphsimp} in order to write $\avjn=j_{\rm tr}$. The assumptions that the particles are spherical and that their radii vary slowly, \ie that their size depends on macroscopic position but neighbouring particles are almost the same size, gives rise to the following relationships
\be \label{geo_cons}
1-\ev(x) = n(x) \frac{4 \pi R(x)^3}{3}, \quad \bet(x)=n(x) 4 \pi R(x)^2
\ee
where $n(x)$ is the number density of electrode particles and, owing to our assumption that volume fraction of binder is negligible, $1-\ev$ is the volume fraction of electrode particles.



{Here we set out the pseudo 2d-model, which follows from the homogenisation described in \S \ref{homog} and describes the performance of a cell at constant temperature. In this model we denote variables and parameters relating to the negative electrode (or anode) by the superscript (a) and variables and parameters relating to the positive electrode (or cathode) by the superscript (c). The transport equations for the electrolyte, obtained from \eqref{presi}-\eqref{dent}, are
\be 
\left. \begin{array}{c}
\ds \ev \frac{\dd c}{\dd t}= \frac{\dd}{\dd x} \left(\de B_{11} \frac{\dd c}{\dd x} - \frac{\tn \avsj}{F} \right)  + \frac{\src}{F},\\*[4mm]
\ds \frac{\dd \avsj}{\dd x} = \src, \\*[4mm]
\ds \avsj=-\kappa(c) B_{11} \left( \frac{\dd \vph}{\dd x} - 2 (1-\tn) \frac{RT}{F} { \frac{a_e'(c)}{a_e(c)} \frac{\dd c}{\dd x} }\right)
\end{array} \right\} \ \mbox{in} \ \ L_1<x<L_4.  \label{pstd-1}
\ee
where the volumetric current source term $\src(x,t)$ is given by
\be 
\src= \left\{ \begin{array}{lll} \ds \bta \jtra(\phisa-\vph,\csa|_{r=\Ra(x)},c), & \mbox{for} & L_1<x<L_2, \\*[4mm]
0, & \mbox{for} & L_2\leq x \leq L_3,\\*[4mm]
\ds \btc \jtrc(\phisc-\vph,\cscc|_{r=\Rc(x)},c), & \mbox{for} & L_3<x<L_4,
\end{array} \right.  \label{pstd-2}
\ee
and where the electrolyte volume fraction $\ev$ and the $B_{11}$ component of the permeability tensor are evaluated appropriately in each of the three regions. As discussed in \S\ref{homog}, $B_{11}$ can be computed by solving the appropriate cell problems. However, a common approach is to instead estimate its value using $B_{11}= \epsilon_v^p$ where $p$ is the Bruggeman porosity exponent (a nondimensional constant), which is commonly taken to be $p=1.5$, see \cite{Brug,gully14} and \cite{gupt}. We note also the work of \cite{shen} who discuss some alternative estimation methods beyond the Bruggeman approximation. Boundary conditions on the electrolyte equations are enforced by the requirements that there is no flow of electrolyte current or flux of lithium ions into the current collectors at $x=L_1$ and $x=L_4$ and are
\be 
\begin{array}{cc}
\avsj|_{x=L_1}=0, & \ds \left. \frac{\dd c}{\dd x} \right|_{x=L_1}=0, \\*[4mm]
\avsj|_{x=L_4}=0, & \ds \left. \frac{\dd c}{\dd x} \right|_{x=L_4}=0.
\end{array}  \label{pstd-3}
\ee
In the negative electrode matrix conservation of current and Ohm's Law, as given by \eqref{ford}, are described by
\be 
\frac{\dd \jsa}{\dd x}= - \src, \ \ \mbox{and} \ \ \jsa=-\kapsa \frac{\dd \phisa}{\dd x}, \quad \mbox{in}  \ \ L_1<x<L_2, \label{pstd-4}
\ee
and are supplemented by a boundary condition at the interface with the current collector, which specifies the current inflow, and one at the interface with the insulating separator into which the current in the matrix does not flow
\be 
\jsa|_{x=L_1}=\frac{I}{A}, \ \ \mbox{and} \ \ \ \jsa|_{x=L_2}=0.  \label{pstd-5}
\ee
Here $I$ is the current flowing into the cell and $A$ is the cell's area. In the same region, lithium transport (as described in \eqref{nixon}-\eqref{spiro}) within the spherical anode particles satisfies the problem
\be 
\frac{\dd \csa}{\dd t} = \frac{1}{r^2} \frac{\dd}{\dd r} \left( r^2 \dsa( \csa) \frac{\dd \csa }{\dd r}  \right), \quad \mbox{for} \quad L_1<x<L_2, \label{pstd-6}\\
\csa \ \ \mbox{bounded on} \ \ r=0, \label{pstd-7}\\
\left. \dsa( \csa) \frac{\dd \csa }{\dd r} \right|_{r=\Ra(x)} =-\frac{\jtra(\phisa-\vph,\csa|_{r=\Ra(x)},c)}{F}. \label{pstd-8}
\ee
In the positive electrode (or cathode) an analogous set of equations and boundary conditions describe the current flow and transport of lithium within the cathode particles. They are
\be 
\frac{\dd \jsc}{\dd x}= - \src, \ \ \mbox{and} \ \ \jsc=-\kapsc \frac{\dd \phisc}{\dd x}, \quad \mbox{in}  \ \ L_3<x<L_4,  \label{pstd-9} \\
\jsc|_{x=L_3}=0, \ \ \mbox{and} \ \ \ \jsc|_{x=L_4}=\frac{I}{A}, \label{pstd-10}
\ee
with lithium transport within the spherical cathode particles being described by
\be 
\frac{\dd \cscc}{\dd t} = \frac{1}{r^2} \frac{\dd}{\dd r} \left( r^2 \dsc( \cscc) \frac{\dd \cscc }{\dd r}  \right), \quad \mbox{for} \quad L_3<x<L_4,\label{pstd-11} \\
\cscc \ \ \mbox{bounded on} \ \ r=0, \label{pstd-12}\\
\left. \dsc( \cscc) \frac{\dd \cscc }{\dd r} \right|_{r=\Rc(x)} =-\frac{\jtrc(\phisc-\vph,\cscc|_{r=\Rc(x)},c)}{F}. \label{pstd-14}
\ee
The transfer currents densities $\jtra(\phisa-\vph,\csa|_{r=\Ra(x)},c)$ and $\jtrc(\phisc-\vph,\cscc|_{r=\Rc(x)},c)$ that describe the flow of current out of the anode and cathode particles, respectively, and which act to couple together the lithium transport problems in the electrolyte to those in the electrode particles are typically determined by the Butler-Volmer condition as discussed in \S \ref{sdbve}. The existence and uniqueness of solution of the system (\ref{pstd-1})-(\ref{pstd-14}) has been proved in \cite{ramos18}.} We point out that, in some works in the literature (see, for instance, \cite{2002GoEtAl}, \cite{2006SmWa}, \cite{2006-b-SmWa}, \cite{2007SmRaWa}, and \cite{2012KiSmIrPe}), authors present the following (incorrect) boundary conditions at $x=L_1$ and $x=L_4$. We note in passing that if these incorrect conditions are used then it can be proved (see \cite{ramos16}) that the corresponding system of boundary value problems does not have any solution unless $I(t) \equiv 0$.

Once the model described above has been solved, we can estimate the state of the charge of the negative electrode SOC$^{(a)}(t)$ and of the positive electrode SOC$^{(c)}(t)$, and the cell voltage $V(t)$, at time $t$, by computing
\begin{equation}
\mbox{SOC}^{(a)}(t)=\frac{3}{(L_2-L_1)(\Ra(x))^3} \int_{L_1}^{L_2}\int_0^{\Ra(x)} r^2 \frac{c^{(a)}_{\rm s}(r,x,t)}{c^{(a)}_{\rm s,max}}{\rm d}r{\rm d}x,
\end{equation}
\begin{equation}
\mbox{SOC}^{(c)}(t)=\frac{3}{(L_4-L_3)(\Rc(x))^3} \int_{L_3}^{L_4} \int_0^{\Rc(x)} r^2 \frac{c^{(c)}_{\rm s}(r,x,t)}{c^{(c)}_{\rm s,max}}{\rm d}r{\rm d}x,
\end{equation}
\begin{equation}
V(t) = \phi_{\rm s} (L_4,t) - \phi_{\rm s} (L_1,t) - \frac{R_{\rm f}}{A}I(t),
\end{equation}
where there is a constant resistance $R_{\rm f}$ which accounts for the potential dropped in the current collectors, but we remark that this is often negligible in practice. We note some ambiguity about how the state of charge of an electrode or cell should be defined. The definition given here might be referred to as the state of charge measured with respect to the theoretical maximum capacity. In the experimental literature it is common to define the state of charge with reference to the capacity measured from a cell under a low-rate (dis)charge, see \eg \cite{johns09,dham}.

\begin{figure}[ht!]
\begin{center}
\includegraphics[width=0.65\columnwidth]{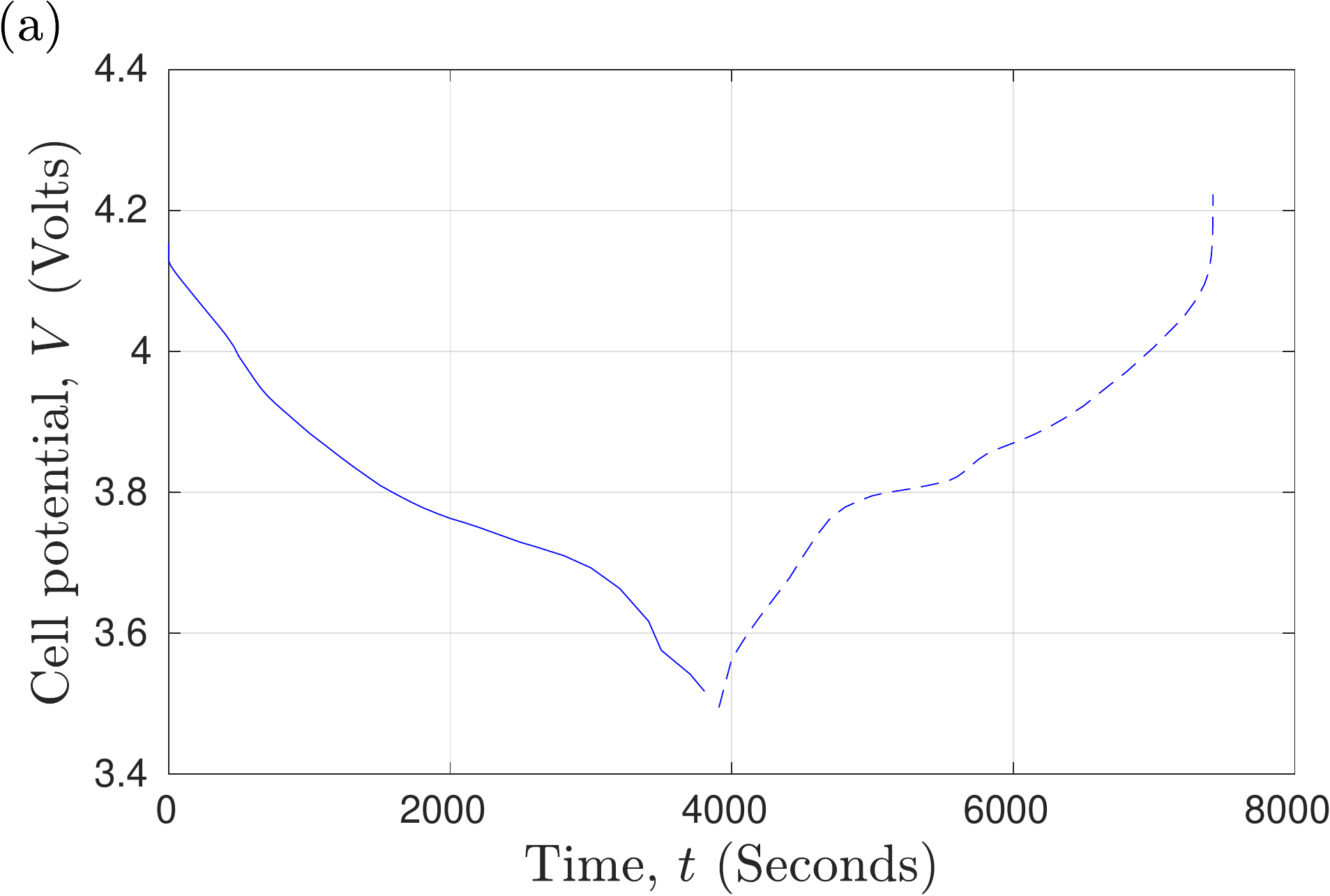}\\
\includegraphics[width=0.49\columnwidth]{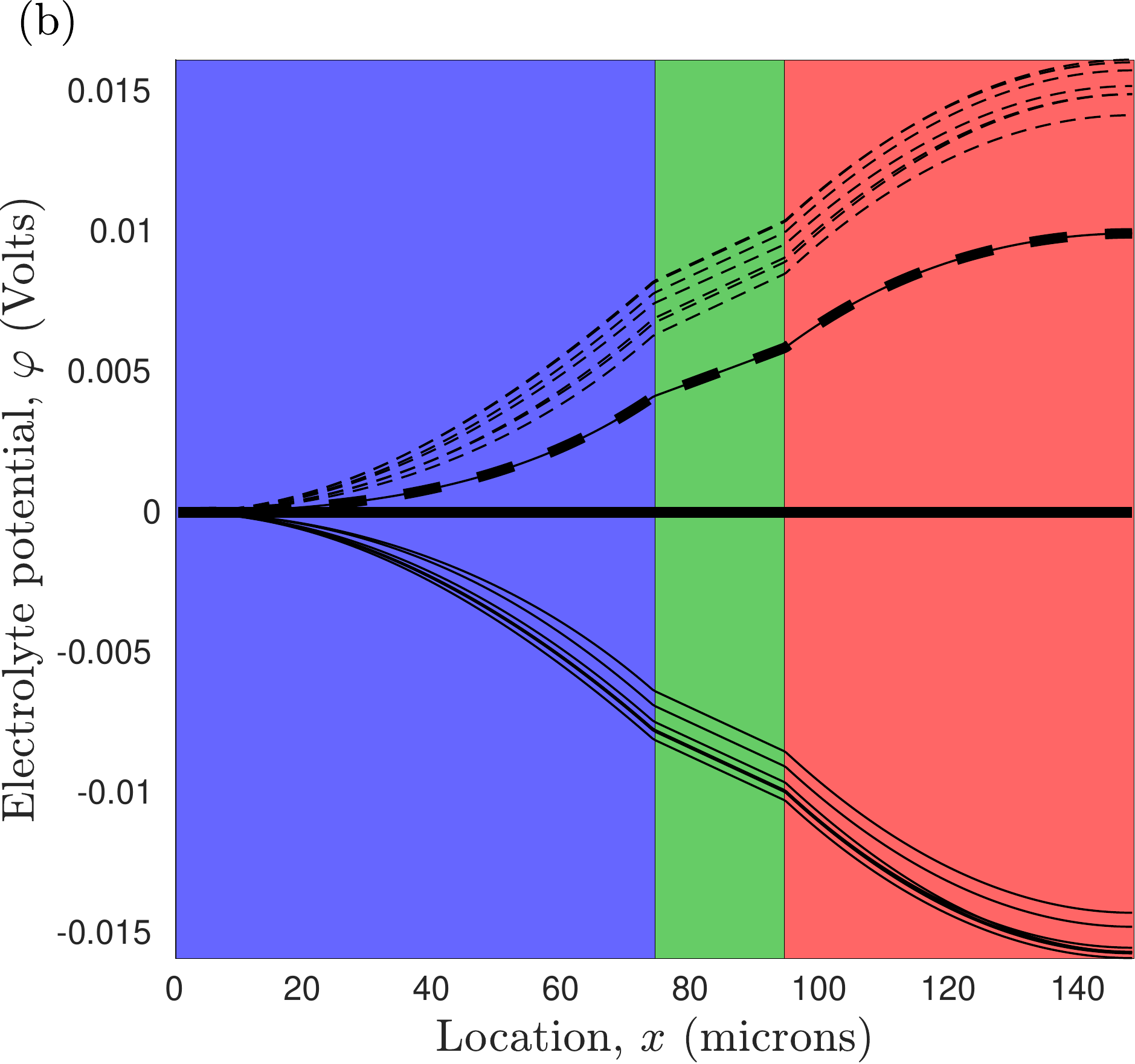}
\includegraphics[width=0.49\columnwidth]{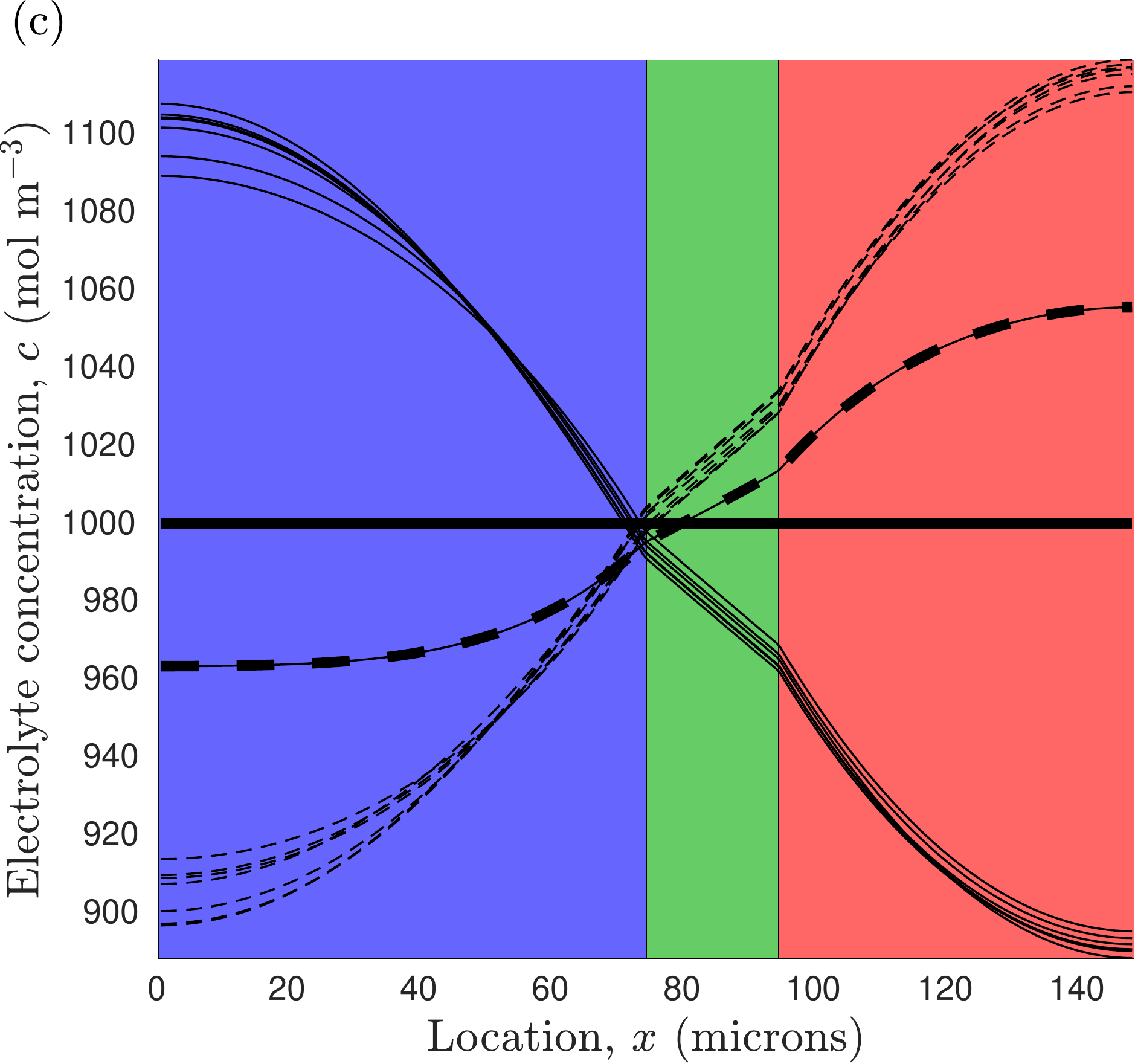}\\
\includegraphics[width=0.49\columnwidth]{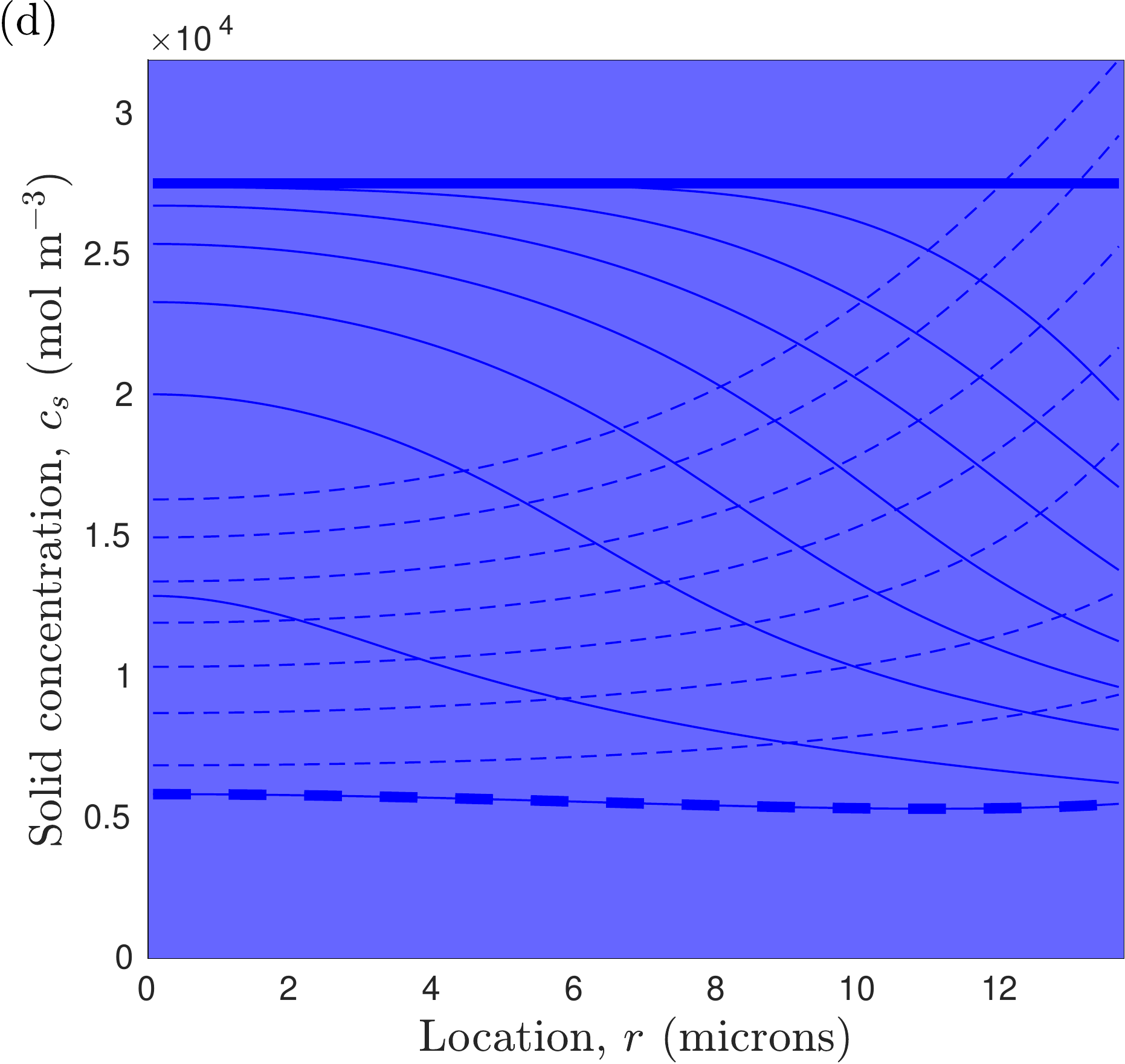}
\includegraphics[width=0.49\columnwidth]{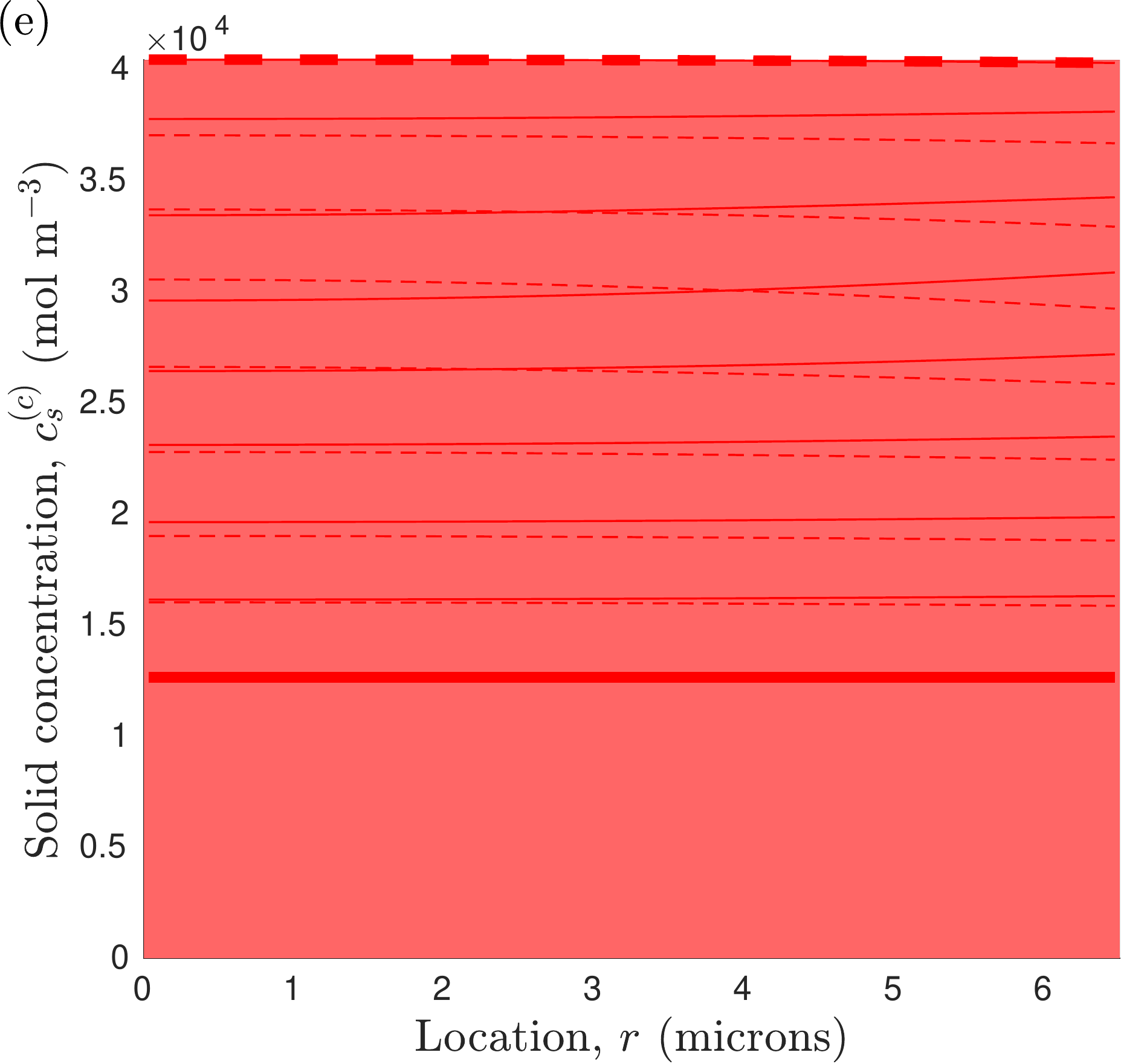}
\caption{\label{fitting_carbon} Discharge (solid curves) and subsequent recharge (dashed curves) of a graphite-LNC cell bathed in 1M LiPF$_6$ electrolyte at a relatively low rate of 0.13A. The full parameterisation is give in Table \ref{simparams}. Thick curves indiciate profiles at the beginning of the (dis)charge stages and the different snapshots are taken every 500s. Panels (a)-(c) show the cell potential, electrolyte potential and electrolyte concentrations respectively, and panels (d) and (e) indicate profiles within an anode and cathode particle both of which are located half way through the thickness of their respective electrodes.} 
\end{center}
\end{figure}

\begin{figure}[ht!]
\begin{center}
\includegraphics[width=0.65\columnwidth]{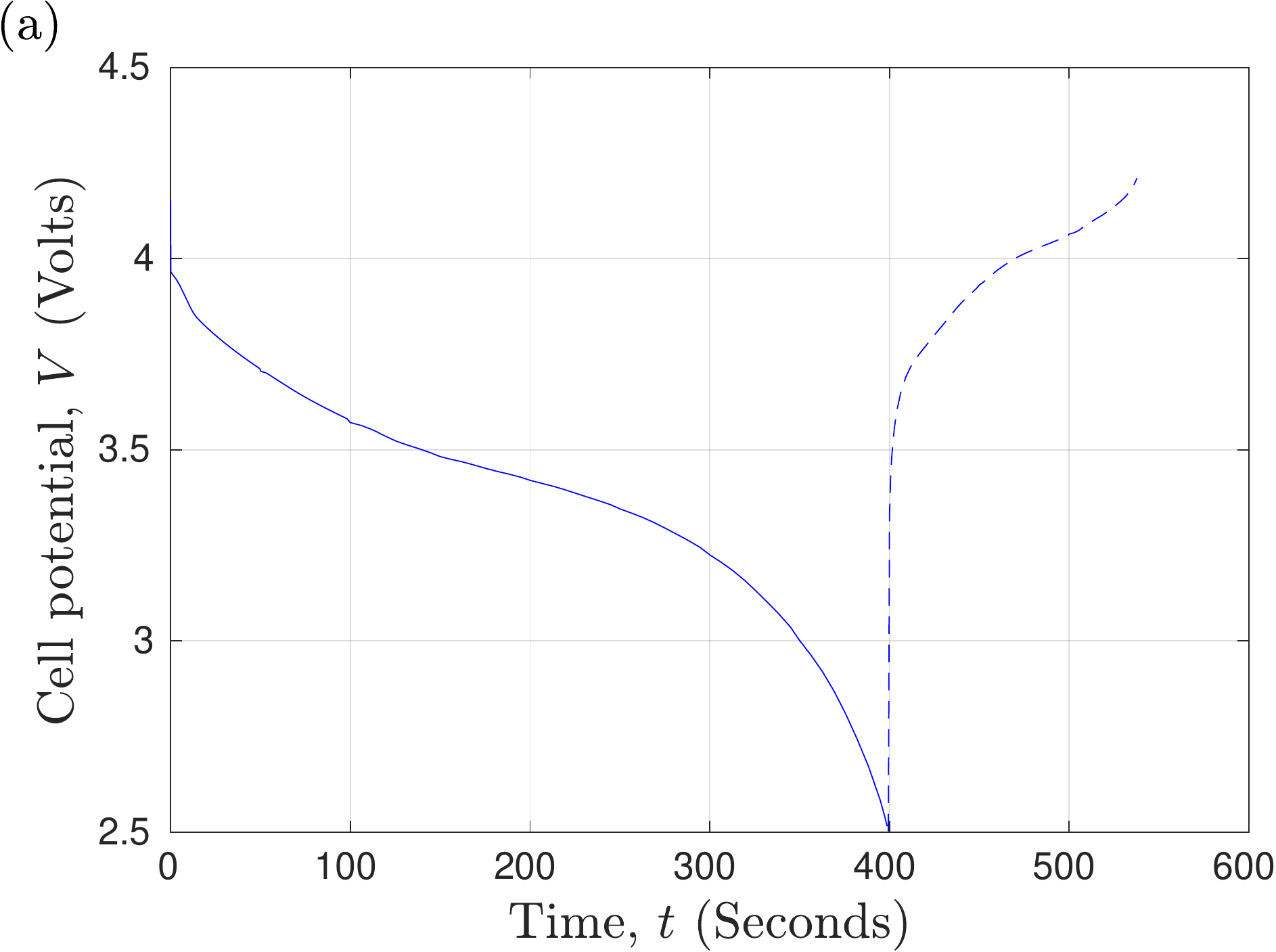}\\
\includegraphics[width=0.49\columnwidth]{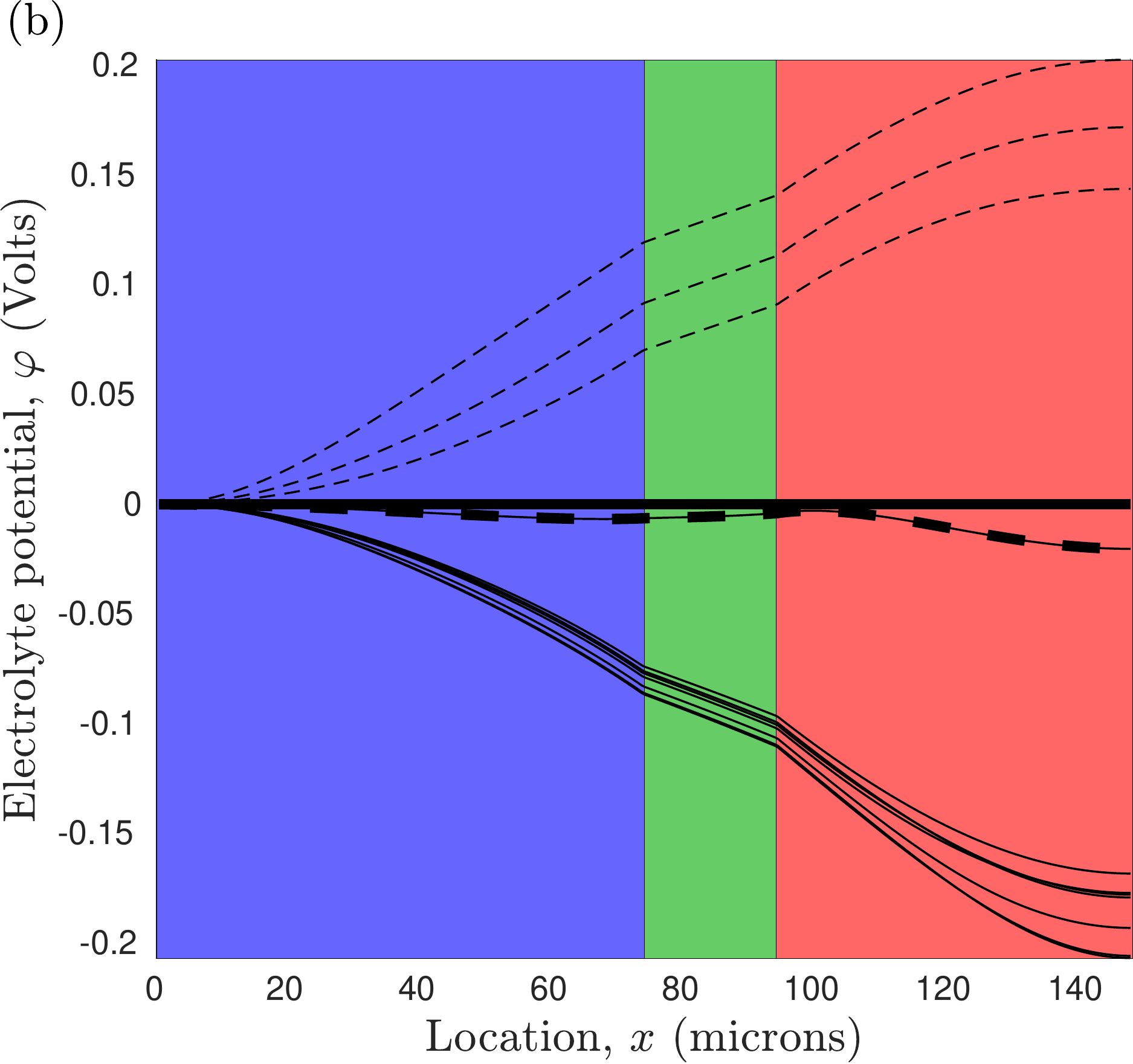}
\includegraphics[width=0.49\columnwidth]{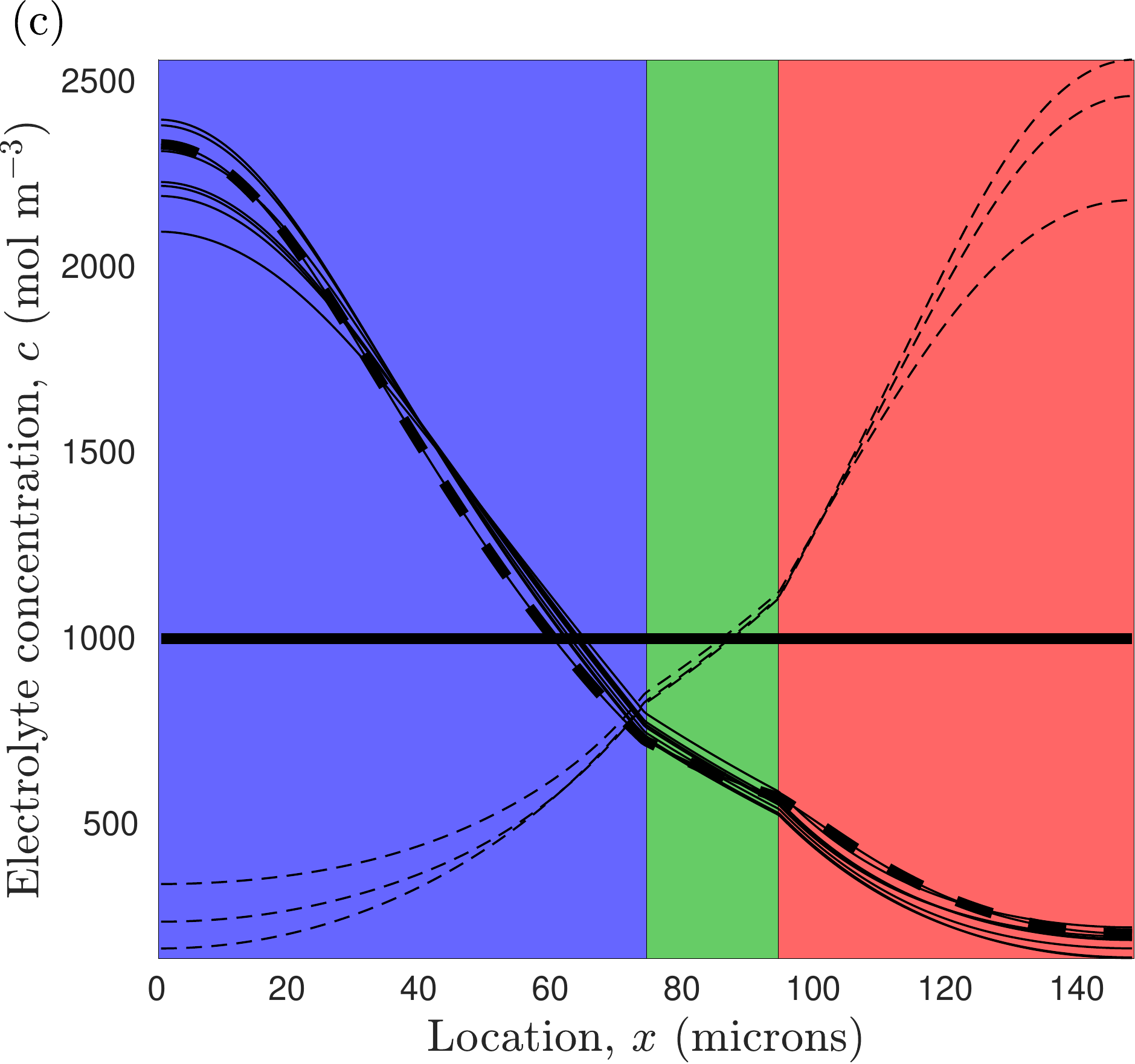}\\
\includegraphics[width=0.49\columnwidth]{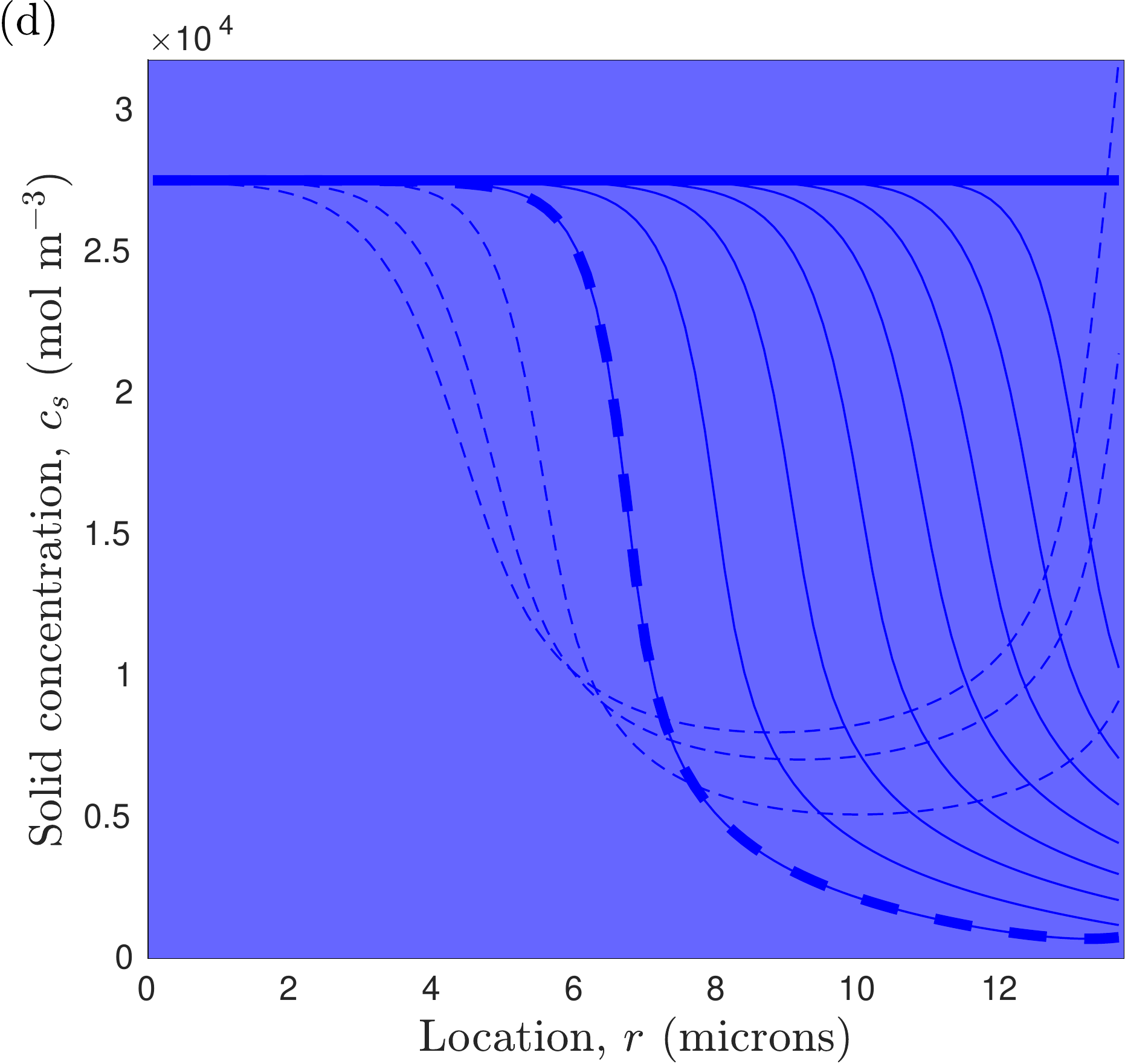}
\includegraphics[width=0.49\columnwidth]{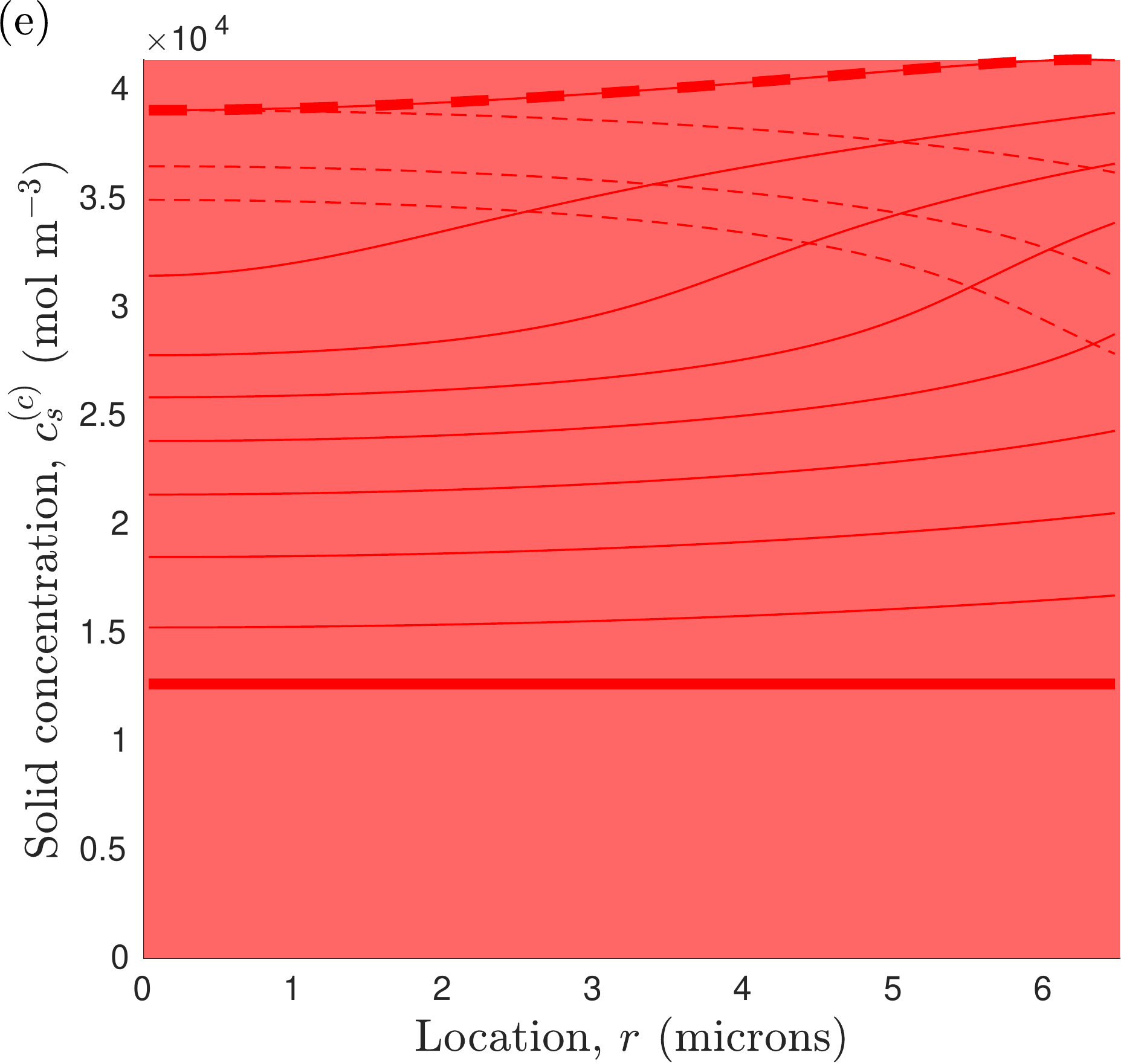}
\caption{\label{fitting_carbon2} Discharge (solid curves) and subsequent recharge (dashed curves) of a graphite-LNC cell bathed in 1M LiPF$_6$ electrolyte at a relatively high rate of 1.3A. The full parameterisation is give in Table \ref{simparams}. Thick curves indiciate profiles at the beginning of the (dis)charge stages and the different snapshots are taken every 50s. Panels (a)-(c) show the cell potential, electrolyte potential and electrolyte concentrations respectively, and panels (d) and (e) indicate profiles within an anode and cathode particle both of which are located half way through the thickness of their respective electrodes.}
\end{center}
\end{figure}


\section{\label{exres}Example results of the model}

To illustrate the capabilities of the model multiscale model, (\ref{pstd-1})-(\ref{pstd-14}), described in \S\ref{p2dsec} we have applied it to model the discharge, and immediate subsequent recharge, of a Li$_x$C$_6$ graphite anode against an Li$_x$(Ni$_{0.4}$Co$_{0.6}$)O$_2$ nickel-cobalt oxide (LNC) cathode which are connected via a 1M LiPF$_6$ in EC:DMC electrolyte. The parameterisation used here is closely based on that given in \cite{EckerI} and \cite{EckerII} where a series of experiments were conducted on a high energy pouch cell produced by Kokam; the values used here are summarised in Table \ref{simparams}. 

Figure \ref{fitting_carbon} shows the discharge curve, and internal concentration and potential profiles during a relatively low-rate usage where a current demand of 0.13A is applied for 4000s and the cell is the immediately recharged at the same rate until it reaches a cut-off voltage of 4.2V. We observe that under these conditions concentration and potential gradients in the electrolyte are relatively modest. Likewise, the concentration is through the radius of the electrode particles is almost uniform; a consequence of the relatively large diffusivity in {LNC}. The largest gradients are observed internal to the anode particles and although these are not large enough to hamper the initial discharge, it is the inability of the graphite to transport intercalated Li from its surface into its interior that ultimately causes the recharging process to be interupted. At the final snapshot in time in panel (d) we observe that the concentration on the surface of the graphite particle has reached its maximum and therefore intercalation cannot proceed further at this location despite their being available space to accomodate Li in the particle's interior. 

Figure \ref{fitting_carbon2} shows the same undergoing a similar discharging and subsequent recharging protocol, but at a more aggresive demand of 1.3A for a shorter time of 400s, followed by a subsequent aggresive recharging again at 1.3A. Note that this faster discharge supplies the same amount of charge to the external circuit as the slower protocol but in a time window 10 times smaller. At the increased rate we observe that gradients in the electrolyte are much more pronounced, and in fact they are sufficiently large that the deep regions of the cathode approach depletion during discharge. The same is true in the anode during recharging. This larger polarisation contributes to a diminished cell voltage during discharge and we can observe that at the deepest discharge state the voltage has dropped to 2.5V, which is markedly lower than the 3.5V attained during the slower protocol despite the devices supplying the same amount of charge. There are now also noticeable concentration gradients within the LNC electrode particles and gradients in the graphite  particles are very high. Once again, it is the graphite which ultimately causes recharging to terminate because the surface of the graphite particles becomes saturated and the recharging can only for around 140s.

\subsection{Numerical solutions to the pseudo 2d-model}
The solutions shown in Figures \ref{fitting_carbon} and \ref{fitting_carbon2} were determined numerically using an in-house ultra-fast and robust solver called {\tt DandeLiion}, the details of which will be described in a forthcoming paper, see \cite{dande}. Whilst is is beyond the scope of this work to reiterate, in detail, the workings of the numerical methods used it is pertinent to outline the approach and highlight some of difficulties in solving (\ref{pstd-1})-(\ref{pstd-14}) numerically. 

First, a spatial mesh is defined. We introduce $N^{(a)}$ grid points across the anode for $x \in (L_1,L_2)$, $N^{(s)}$ across the separator for $x \in (L_2,L_3)$ and $N^{(c)}$ across the cathode for $x \in (L_3,L_4)$. At each point in the anode and cathode a microscopic transport problem must be solved and so at each of the $N^{(a)}+N^{(c)}$ grid points a further $M$ grid points need to be introduced to on which to discretise the microscopic transport equations within the electrode particles. Consequently, the complete discrete geometry is comprised of $(N^{(a)}+N^{(c)}) \times M + N^{(a)} + N^{(s)} + N^{(c)}$ grid points. Second, a suitable approximation (\eg finite volumes or finite elements) can be used to remove the spatial derivatives and reduce the problem to a large system of coupled differential-algebraic equations (DAEs). The algebraic equations arise largely from the elliptic PDEs, \eg those for the electron conduction in the solid (\ref{pstd-4}), whereas the ordinary differential equations arise from the parabolic PDEs, \eg those for transport in the electrode particles (\ref{pstd-6})-(\ref{pstd-8}). We note the importance of using a spatial discretisation method which is conservative; if such a method is not used, on repeated cell cycling,  the total amount of lithium within the system changes markedly and introduces significant errors. It is for this reason that many approaches based on finite difference approximations are not recommended. Third, a scheme for timestepping the system of DAEs must be found. The {\tt DandeLiion} software uses uses a selection of implicit backward differentiation formula methods, of orders 1-6, and also offers adaptive time stepping. The choices of timestepping methods is restricted, in comparison to those that can be used for pure ODE systems, because of the additional constraints imposed by the algebraic equations. 

Implementing the steps outlined above and implementing in {\tt C++} gives rise to a numerical scheme for solving the pseudo 2d-model in a very short time on a standard desktop computer. For reference, the simulation results shown in Figures \ref{fitting_carbon} and \ref{fitting_carbon2} each took around 5 seconds to run and were performed on a discretised geometry comprised of 20,300 grid points with 100 spatial points inside the anode, separator, cathode and each of the electrode particles.


\begin{table}
\begin{center}
\caption{\label{simparams} The parameter values used to carry out the simulations shown in \S\ref{exres}. These are largely based on the work of \cite{EckerI} and \cite{EckerII}. The functions used for the electrode conductivities were fitted to data in \cite{EckerI} and \cite{EckerII} and the functions themselves are given in \cite{dande}.}
\begin{tabular}{ c|c|c|c } 
\hline
{\bf Type} & {\bf Parameter} & {\bf Symbol} & {\bf Units}  \\
\hline
{\bf Anode} 		&		Thickness 				& $L_4-L_3$ 	& 74$\times$10$^{-6}$m 	 \\
			&		Volume fraction of electrolyte		& $\ev$		& 0.329			 \\
			&		Permeability tensor component		& $B_{11}$	& 0.162			 \\
			&		BET surface area			& $\bta$	& 81548m$^{-1}$		 \\
			&		Particle radius				& $\Ra$		& 13.7$\times$10$^{-6}$m	 \\
			&		Electrode conductivity			& $\kapsa$	& 14 S m$^{-1}$ \\
			&		Diffusivity in anode particles		& $\dsa(c^{(a)}_s)$		& See caption	\\
			&		Maximum concentation of Li in anode particles &	$c^{(a)}_{\rm s,max}$	& 31920 mol m$^{-3}$	\\
			&		Initial concentration of Li in anode particles & $c^{(a)}_{\rm s}|_{t=0}$	& 27523 mol m$^{-3}$	\\
\hline
{\bf Cathode} 		&		Thickness 				& $L_2-L_1$ 	& 54$\times$10$^{-6}$m  \\
			&		Volume fraction of electrolyte		& $\ev$		& 0.296			\\
			&		Permeability tensor component		& $B_{11}$	& 0.1526		\\
			&		BET surface area			& $\btc$	& 188455m$^{-1}$	\\
			&		Particle radius				& $\Rc$		& 6.5$\times$10$^{-6}$m  \\
			&		Electrode conductivity			& $\kapsc$	& 68.1 S m$^{-1}$	\\
			&		Diffusivity in cathode particles	& $\dsc(c^{(c)}_s)$	& See caption	\\
			&		Maximum concentation of Li in cathode particles & $c^{(c)}_{\rm s,max}$	& 48580 mol m$^{-3}$	\\
			&		Initial concentration of Li in cathode particles & $c^{(c)}_{\rm s}|_{t=0}$	& 12631 mol m$^{-3}$ \\
\hline
{\bf Separator} 	&		Thickness				& $L_3-L_2$ 	& 20$\times$10$^{-6}$m  \\
			&		Volume fraction of electrolyte		& $\ev$		& 0.508			\\
			&		Permeability tensor component		& $B_{11}$	& 0.304			\\
\hline
{\bf Electrolyte} 	&		{Transference number} 		& $\tn$		& 0.26 \\
			&		Conductivity				& $\kappa(c)$	& See Figure \ref{fitting} 	\\
			&		Diffusivity				& $\de(c)$	& See Figure \ref{fitting} 	\\
			&		Initial concentration			& $c|_{t=0}$	& 1000 mol m$^{-3}$	\\
\hline
{\bf Global} 		&		Area				 	& $A$ 		& 8.585$\times$10$^{-3}$m$^2$ \\
			&		Film resistance				& $R_{\rm f}$	& 0 $\Omega$ \\
			& 		Temperature				& $T$		& 298.15 K \\
\hline
\end{tabular}
\end{center}
\end{table}



\section{Conclusion \label{conclusions}}

We have reviewed the existing modelling of charge transport models of lithium ion batteries at the cell scale. This includes a description of ionic motion in the electrolyte in both the dilute and the more practically relevant moderately-concentrated regimes. We resolve a common source of confusion in electrolyte modelling in the literature which arises from the definition of the electric potential. In the electrochemical literature this is usually chosen to be the potential measured with respect to metallic lithium electrode, in contrast the standard definition used in the physics community it is with respect to a vacuum at infinity. Crucially this choice of potential affects the coefficients in the electrolyte transport equations.  We
have also examined the Butler-Volmer relation describing reaction at
the interface between the electrolyte and the solid electrode
particles and demonstrated that these should have a particular
functional form in order to avoid nonphysical predictions. The
dependency in the Butler-Volmer relation should not only account for
the solid electrode particle becoming completely intercalated or
deintercalated but allow for cases where the electrolyte concentration
gets very low. The specific behaviour of various common solid materials
used in electrodes have been considered including the dominant
mechanisms for lithium transport and the possible modelling approaches
that can be used. The problem of up-scaling the models from the
microscopic (a single electrode particle) to the macroscale (the
whole cell) has been considered and the appropriate approximations
discussed that allow the models to account for moderately-concentrated
electrolyte behaviour reviewed. In addition numerical solution to the Newman model is discussed and some representative realistic solutions to the resulting pseudo 2-dimensional model have been presented and discussed. It is our hope that this work will prove a useful guide to people who are new to this topic allowing them to develop an appreciation of this highly fertile and technologically important area of research.

\section{Acknowledgements}
{{GWR, JMF and CPP are supported by the Faraday Institution Multi-Scale Modelling (MSM) project Grant number EP/S003053/1.} AMR gives thanks for the financial support of the Spanish “Ministry of Economy and Competitiveness” under project MTM2015-64865-P for this work.}

\bibliographystyle{agsm} 
\bibliography{batterybiblio}

\appendix

\begin{table}
\begin{center}
\caption{\label{nomen} A summary of the most important nomenclature.}
\begin{tabular}{ c|c|c|c } 
\hline
{\bf Type} & {\bf Parameter} & {\bf Symbol} & {\bf Units} \\
\hline
{\bf Geometrical} &		Anode thickness 	& $L_2-L_1$ 	& m \\
&				Separator thickness 	& $L_3-L_2$ 	& m \\
&				Cathode thickness	& $L_4-L_3$ 	& m \\
&				BET surface area	& $\bet$	& 1/m \\
&				Permeability factor	& $\B$		& none \\
&				Volume fraction of electrolyte & $\ev$	& none \\
&				Electrode particle radius & $R$		& m \\
\hline
{\bf Electrolyte} 	& 	Ionic concentration  	& $c$ 			& mol/m$^3$ 	\\
&				Total molar concentration & $c_T$		& mol/m$^3$ 	\\
&				Electric potential  	& $\phi$ 		& V 		\\
&				Electric potential measured with respect to a Li electrode  	& $\vph$ 		& V \\
&				Current density  	& $j$ 			& A/m$^2$ \\
&				Electric field		& $\vect{E}$ 		& V/m \\
&				Permittivity		& $\varepsilon$		& F/ms	\\
&				Molar concentration of positive and negative ions & $c_p$ and $c_n$ & mol/m$^3$ \\
&				Molar flux of positive and negative ions & $\qp$ and $\qn$ & mol/m$^2$s \\
&				Average velocities of positive and negative ions & $\vn$ and $\vp$ & m/s \\
&				\makecell{Component of the average velocity of positive and\\negative ions due to the electric field} & $\vect{v}_{ep}$ and $\vect{v}_{en}$ & m/s \\
&				Diffusivities of positive and negative ions & $D_p$ and $D_n$ & m$^2$/s \\
&				Ionic diffusion coefficient	& $\dc$ & m$^2$/s \\
&				Mobilities of positive and negative ions & $M_p$ and $M_n$ & m$^2$/Vs \\
& 				Electrochemical potential of positive and negative ions & ${\mu}_n$ and ${\mu}_p$ & J/mol \\
& 				Chemical potential of positive and negative ions & $\bar{\mu}_n$ and $\bar{\mu}_p$ & J/mol \\
& 				\makecell{Standard state potential of\\positive and negative ions} & $\mu^0_n$ and $\mu^0_p$ & J/mol \\
&				Effective ionic diffusivity & $\de$ & m$^2$/s \\
&				Conductivity (dilute theory) & $\hat{\kappa}$ & S/m \\
&				Conductivity (moderately concentrated theory) & $\kappa$ & S/m \\
&				Transference number & $t_+$ & none \\
&				\makecell{Transference number of positive ions\\with respect to the solvent velocity} & $t^0_+$ & none \\
\hline
{\bf Electrode} 	&	Electric potential 	& $\phi_s$		& V		\\
{\bf matrix}		&	Current density		& $j_s$			& A/m$^2$	\\
&				Effective conductivity	& $\kappa_s$		& S/m		\\
\hline
{\bf Electrode}		&	Li concentration	& $c_s$			& mol/m$^3$	\\
{\bf particles}		&	Diffusion coefficient	& $D_s$			& m$^2$/s	\\
&				Maximum concentration of Li & $c_{\rm s,max}$   & mol/m$^3$	\\
\hline
{\bf Interfacial} &		Transfer current density & $j_{\rm tr}$		& A/m$^2$	\\
&				Open circuit potential & $U_{\rm eq}$		& $V$		\\
&				Exchange current density & $i_0$		& A/m$^2$	\\
&				Anodic and cathodic transfer coefficients & $\alpha_{\rm a}$ and $\alpha_{\rm c}$ & none \\
&				Contact film resistance	& $R_{\mbox{f}}$	& $\Omega$ \\
\hline
\end{tabular}
\end{center}
\end{table}

\section{The chemical potentials of a mixture and the Gibbs-Duhem Relation \label{appa}}
Consider a homogeneous mixture containing $K$ different species with mole fractions $\chi_1, \chi_2, \cdots, \chi_K$. The chemical potentials of this mixture $\mub_1, \mub_2, \cdots, \mub_K$ are defined, in terms of the Gibbs free energy $G$, by the relations
\be
\mub_i= \frac{\dd G}{\dd n_i} \qquad \mbox{for} \quad i=1,2, \cdots,K,  \label{chempot}
\ee
where $n_i$ is the number of moles of species $i$ in the mixture. The Gibb's free energy of the system clearly scales linearly with the total number of moles of the mixture $N_{tot}$ when the mole fractions of the various species are held constant. It follows that we can write
\be
G(N_{tot},\chi_1, \chi_2, \cdots, \chi_K)=N_{tot} h(\chi_1, \chi_2, \cdots, \chi_K), \label{Gdef}
\ee
where here $h(\cdot)$ is the Gibbs free energy of the mixture per mole. Given that $N_{tot}$ and the various mole fractions are defined in terms of the number of moles of each species in the mix by
\be
N_{tot}=\sum_{k=1}^K n_k \qquad \chi_i=\frac{n_i}{\sum_{k=1}^K n_k}
\ee
we can use the chain rule to re-express the derivatives in (\ref{chempot}) in the form
\bes
 \frac{\dd G}{\dd n_i}&=&\frac{\dd G}{\dd N_{tot}}\frac{\dd N_{tot}}{\dd n_i}+  \sum_{j=1}^K  \frac{\dd G}{\dd \chi_j}\frac{\dd \chi_j}{\dd n_i} \non \\
 &=&\frac{\dd G}{\dd N_{tot}}+  \sum_{j=1}^K  \frac{\delta_{ij} -\chi_j}{N_{tot}}\frac{\dd G}{\dd \chi_j}.
\ees
It follows on substituting for $G$ from (\ref{Gdef})  that the chemical potentials are given in terms of $h$ by
\be
\mub_i(\chi_1, \chi_2, \cdots, \chi_K)=h+\frac{\dd h}{\dd \chi_i} - \sum_{j=1}^K  \chi_j \frac{\dd h}{\dd \chi_j}  \label{chempot2}
\ee
Thus the chemical potentials of the mixture are functions only of the mole fractions of the various species that it contains. The relation (\ref{chempot2}) taken together with the obvious relation between the mole fractions
\be
\sum_{k=1}^K \chi_k=1 \label{obvious}
\ee
leads to a further relationship between the chemical potentials, the Gibbs-Duhem relation. We derive this by taking the derivative of (\ref{chempot2}) with respect to the mole fraction $\chi_p$ to obtain
\bes
\frac{\dd \mub_i}{\dd \chi_p}= \frac{\dd^2 h}{\dd \chi_i \dd \chi_p} -\sum_{k=1}^K\chi_j \frac{\dd^2 h}{\dd \chi_j \dd \chi_p}.
\ees
Multiplying this equation by $\chi_i$ and summing the result between $i=1$ and $i=K$ gives
\bes
\sum_{i=1}^K \chi_i \frac{\dd \mub_i}{\dd \chi_p}= \left(\sum_{i=1}^K \chi_i \frac{\dd^2 h}{\dd \chi_i \dd \chi_p}\right) -\left( \sum_{i=1}^K \chi_i \right)\left( \sum_{k=1}^K\chi_j \frac{\dd^2 h}{\dd \chi_j \dd \chi_p} \right),
\ees
which simplifies, on application of (\ref{obvious}), to a version of the Gibbs-Duhem equation
\be
\sum_{i=1}^K \chi_i \frac{\dd \mub_i}{\dd \chi_p}=0.~~~~~~~~~~~~~~~~~~ \label{GD2}
\ee

\end{document}